\documentclass[pdflatex,sn-mathphys-num]{sn-jnl}

\usepackage[utf8]{inputenc}
\usepackage[T1]{fontenc}
\usepackage{rotating}
\usepackage{graphicx}%
\usepackage{multirow}%
\usepackage{amsmath,amssymb,amsfonts}%
\usepackage[title]{appendix}%
\usepackage{xcolor}%
\usepackage{textcomp}%
\usepackage{manyfoot}%
\usepackage{wasysym}
\usepackage{booktabs}%
\usepackage{newunicodechar}
\newunicodechar{ }{\,}
\usepackage{algorithm}%
\usepackage{algorithmicx}%
\usepackage{algpseudocode}%
\usepackage{adjustbox}
\usepackage{adjustbox}
\usepackage{hyperref}
\usepackage{mathrsfs}
\usepackage{listings}%
\usepackage{array}
\usepackage{amsmath}
\usepackage{adjustbox}
\usepackage{geometry}
\usepackage{pdflscape}
\usepackage{lscape}
\usepackage[font=normalsize]{caption}
\usepackage{wasysym}

\newgeometry{vmargin={20mm}, hmargin={27mm,27mm}}

\theoremstyle{thmstyleone}%
\theoremstyle{thmstyletwo}%

\theoremstyle{thmstylethree}%

\begin{document}
	
	\title[Article Title]{2D MXene-Based Photocatalysts for Efficient Water Splitting and Hydrogen Evolution: A brief review}
	
	
	\author[1]{\fnm{C. B.} \sur{Subba}}
	\equalcont{These authors contributed equally to this work.}
	
	\author[2]{\fnm{Prasad} \sur{Mattipally}}
	\equalcont{These authors contributed equally to this work.}
	
	\author[3]{\fnm{J} \sur{Sivakumar}}
	\equalcont{These authors contributed equally to this work.}
	
	\author[4]{\fnm{A.} \sur{Laref}}
	\equalcont{These authors contributed equally to this work.}
	
	\author[5]{\fnm{A.} \sur{Yvaz}}
	\equalcont{These authors contributed equally to this work.}
	
	\author*[1]{\fnm{D. P.} \sur{Rai}} \email{dibyaprakashrai@gmail.com}
	\equalcont{These authors contributed equally to this work.}
	
	\author[1]{\fnm{Z.} \sur{Pachuau}}
	\equalcont{These authors contributed equally to this work.}
	
	\affil*[1]{\orgdiv{Department of Physics}, \orgname{Mizoram University}, \orgaddress{\city{Aizawl}, \postcode{796004}, \country{India}}}
	
	\affil[2]{\orgdiv{Department of Physics}, \orgname{Hyderabad Institute of Technology and  Management (HITAM)}, \orgaddress{ \street{Gowdavelly Medchal-Malkajigiri}, \city{ Hyderabad}, \postcode{501401}, \country{India}}}
	
	\affil[3]{\orgdiv{Department of Physics}, \orgname{JNTUH UCEST}, \orgaddress{\city{Kukatpally, Hyderabad}, \country{India}}}
	
	\affil[4]{\orgdiv{Department of Physics and Astronomy}, \orgname{College of Science, King Saud University}, \orgaddress{\city{Riyadh}, \postcode{11451}, \country{Saudi Arabia}}}
	
	\affil[5]{\orgdiv{World-class research center "Advanced Digital Technologies"}, \orgname{State Marine Technical University}, \orgaddress{\city{Saint Petersburg}, \postcode{190121}, \country{Russia}}}
	
	
	\abstract{Photocatalytic water splitting offers a viable and sustainable method for hydrogen production. MXenes, a class of 2D transition-metal carbides/nitrides, have emerged as potential photocatalysts and cocatalysts due to their tunable electronic properties, high conductivity, and surface functionality. This review explores recent advances in MXene-based photocatalysts for hydrogen production, discussing their synthesis, electronic structures, and photocatalytic mechanisms. The key challenges, including stability issues, charge recombination, and bandgap optimisation, are critically analysed. Finally, future research directions are outlined to improve MXene-based systems for large-scale hydrogen production.}

	
	
	
	\maketitle
	
	\section{Introduction}\label{sec1}
	
	Photocatalytic water splitting is initiated by the absorption of photons, leading to the formation of electron-hole pairs within the material's bandgap. Subsequently, these charge carriers migrate to the catalyst surface, where they can drive redox reactions with water, provided that the potential difference exceeds the 1.23~eV threshold, as established in the foundational work\cite{fujishima1972electrochemical}. The first reported instance of water splitting into oxygen and hydrogen was documented in 1972 \cite{fujishima1972electrochemical}, demonstrating that visible light alone could drive the reaction without the need for an external voltage. UV-light-induced water splitting using powdered \(\mathrm{TiO_2}\) and \(\mathrm{SrTiO_3}\) photocatalysts was first documented in 1980 \cite{sato1980photodecomposition,schrauzer2002photolysis,lehn1980photochemical}.

	Since then, much research has been done using \(\mathrm{TiO_2}\) as a photocatalyst from water decomposition to environmentally remediation\cite{1572261549618605952,NAKATO199535,doi:10.1021/jp962992u,Stalder_1979,KAVAN1995643,Hattori_2000,meng2019dual,lee2013tio2,park2013surface,ochiai2012photoelectrochemical,sayama2001stoichiometric,vinu2010environmental,purabgola2022graphene,ochiai2012photoelectrochemical,kato2002visible,momeni2015visible}.
	Most of the research in photocatalysis has focused primarily on \(\mathrm{TiO_2}\), with the maximum research efforts focused solely on this material to the point that \(\mathrm{TiO_2}\) has often been considered as a prototype of transition metal oxide surfaces \cite{diebold2003surface,pang2008chemical,hammer2010water,dohnalek2010thermally}.
	
	Recently, a quantum efficiency approaching unity has been achieved \cite{takata2020photocatalytic}.
	However, TiO$_2$ based particulate photocatalysts primarily utilise ultraviolet (UV) light for photocatalysis,  
	but the infrared (IR) and visible region, which together carry almost 96\% of the total solar energy\cite{gueymard1995smarts2}, is more desirable 
	for efficient solar energy conversion.
	
	Recently, Du \textit{et al}.~\cite{du2025covalent} reported a nonstoichiometric $\beta$-ketoenamine-linked covalent organic framework capable of photocatalytic activity without the use of cocatalysts. Under irradiation with visible light and utilizing ascorbic acid as a sacrificial agent, the framework exhibited hydrogen evolution rates of 15{,}480~µmol·g$^{-1}$·h$^{-1}$ from seawater and 22{,}450~µmol·g$^{-1}$·h$^{-1}$ from pure water.
	To the best of our knowledge, this is the most efficient photocatalytic hydrogen production without co-catalyst reported to date, setting a new benchmark for future advancements in the field.
	
	Building on this remarkable achievement and accumulated knowledge, it is imperative to explore new materials that can efficiently harness the higher-wavelength regions of the solar spectrum (infrared and visible), while also being abundant and sustainable, in order to enhance overall efficiency and maximize solar energy conversion.
	
	Since the groundbreaking discovery of the two-dimensional (2D) nanomaterial MXenes by Michel W. Barsoum and Yury Gogotsi along with their research team at Drexel University, Philadelphia, in 2011\cite{https://doi.org/10.1002/adma.201102306}, extensive research has been conducted to explore its physical\cite{fatima2023structural,champagne2020physical}, electrical\cite{fatima2023structural,feng2017structures}, mechanical\cite{kazemi2019super,kurtoglu2012first,ren2024avian}, and chemical\cite{li2022mxene,lei2015recent} attributes. MXene exhibits remarkable properties, making it suitable for diverse applications such as energy storage devices (batteries and supercapacitors)\cite{lukatskaya2013cation,ali2023ti3sic2}, sensors\cite{bhardwaj2021mxene,reddy2024recent,wang2021high,xu2020latest}, biomedical tools\cite{li2024recent,seidi2023mxenes}, thermal-to-electrical energy conversion\cite{wang2022mxene}, electromagnetic shielding\cite{verma2023review,iqbal20202d}, catalysis\cite{liu2018termination,reghunath2021synthesis} and superconductor\cite{kamysbayev2020covalent,sevik2023superconductivity}. Since 2014 the attention of researchers has been towards MXene as a potential photocatalyst/co-catalyst and also been investigated as a photocatalyst for the degradation of environmental pollutants in both air and water.
	
	MXenes exhibit superior properties compared to other existing 2D materials, making them highly promising for various technological applications. Their unique combination of metallic conductivity, hydrophilicity, tunable surface chemistry, and structural versatility provides significant advantages over materials such as graphene, transition metal dichalcogenides (TMDs), and hexagonal boron nitride (h-BN). These characteristics enable MXenes to outperform other two-dimensional materials in applications including energy storage, electromagnetic interference shielding, catalysis, and biomedical technologies~\cite{gogotsi2023future,jayakumar20182d}.
	
	When compared to other photocatalytic materials, including oxide-based materials available in previous reviews\cite{kudo2009heterogeneous,maeda2010photocatalytic,hisatomi2014recent,maeda2011photocatalytic} and metal chalcogenides\cite{wolff2018all,ran2014earth,ellis1977study}, transition metal dichalcogenides (TMDs)\cite{li2021carbon}, transition metal oxides (TMOs)\cite{das2022transition}, clay-based materials\cite{ruiz2023mxenes,fatimah2022clay}, graphene\cite{gautam2023review,jin2018superior}, layered double hydroxides (LDH)\cite{lin2023research}, and metal oxide/sulfide-based photocatalysts such as molybdenum sulfide (\(\mathrm{MoS_x}\)), molybdenum disulfide (\(\mathrm{MoS_2}\))\cite{das2021one}, tungsten disulfide (\(\mathrm{WS_2}\))\cite{matte2010mos2}, cerium dioxide (\(\mathrm{CeO_2}\))\cite{liu2022ceo2,wang2023room}, bismuth sulfide (\(\mathrm{BiS_x}\))\cite{yang2020synergistic}, cadmium sulfide (\(\mathrm{CdS}\))\cite{zheng2023recent}, zinc sulfide (\(\mathrm{ZnS}\))\cite{rasool2024interfacial}, vanadium pentoxide (\(\mathrm{V_2O_5}\))\cite{xu2021three}, silicon nitride (\(\mathrm{SiN_x}\))\cite{wozniak2020influence}, zinc oxide (\(\mathrm{ZnO}\))\cite{guo2023cold,yuan2023situ}, titanium dioxide (\(\mathrm{TiO_2}\))\cite{wang2020humidity}, niobium pentoxide (\(\mathrm{Nb_2O_5}\))\cite{zhang2016synthesis}, tantalum pentoxide (\(\mathrm{Ta_2O_5}\))\cite{rafieerad2021development}, tungsten trioxide (\(\mathrm{WO_3}\))\cite{warsi2022synthesis}, and strontium titanate (\(\mathrm{SrTiO_3}\))\cite{meng2024synergistic}, MXene demonstrates superior photocatalytic efficiency\cite{zhong2021two,hong2020recent,tang2020ti3c2}. This is attributed to its large surface-to-volume ratio, outstanding electrical and optical characteristics, and diverse chemical compositions with tunable functional groups, which are obtained via wet chemical etching\cite{murali2022review,gogotsi2023future}.
	
	MXene also features a tunable bandgap, which is influenced by different functional groups \cite{naguib2021ten}. This unique ability allows for fine-tuning its photocatalytic activity to enhance performance significantly\cite{you2021state}. Consequently, it exhibits superior potential compared to its counterparts in photocatalytic applications \cite{yao20202d, biswal2021recent, xie2020positioning}.
	
	Unlike conventional 2D materials, MXene possesses several advantageous properties, such as abundant active sites\cite{he2021dynamically,yuan2018mxene}, an exceptional Young’s modulus\cite{firestein2020young,lipton2019mechanically}, remarkable thermal and chemical stability\cite{zhang2018chemical,li2015synthesis}, an expanded inter-layer distance\cite{xu2021mxene}, active functional groups\cite{hu2018surface}, and high thermal and electrical conductivity\cite{liu2018high}. These attributes collectively establish MXene as a robust and highly efficient photocatalyst, outperforming traditional materials that lack such extraordinary physicochemical properties\cite{yao20202d,biswal2021recent,xie2020positioning}.
	Transition elements, also known as transition metals, possess partially filled d orbitals and play a crucial role in photoexcitation and photocatalytic activity observed~\cite{zou2001direct}.
	
	This review provides a comprehensive framework for systematically understanding MXene, a highly promising 2D nanomaterial, and its potential as a photocatalyst and co-catalyst for water splitting under sunlight. It particularly emphasizes MXene's effectiveness in hydrogen generation—an aspect often overlooked in previous reviews \cite{solangi2023mxene,KUANG202018,nath2016environment,zhong2021two,biswal2022review,kudo2009heterogeneous,bignozzi2013nanostructured,wang2014semiconductor,li2022advances,zhang2018photocatalytic,li2021applications,sajid2024recent,wang2019particulate,ajmal2025advancements,haneef2022recent,twilton2017merger,bai2021recent,praus20242d,moniz2015visible,zarei2017fundamentals}, which did not consider MXene as a viable photocatalyst for green hydrogen generation. Additionally, the review discusses the optimal synthesis routes for achieving higher efficiency. 
	
	In conclusion, the authors analyzed the future challenges and potential advancements of MXene-based photocatalysts for water splitting in hydrogen production.
	\begin{figure*}
		\centering
		\fbox{\includegraphics[width=1.0\textwidth]{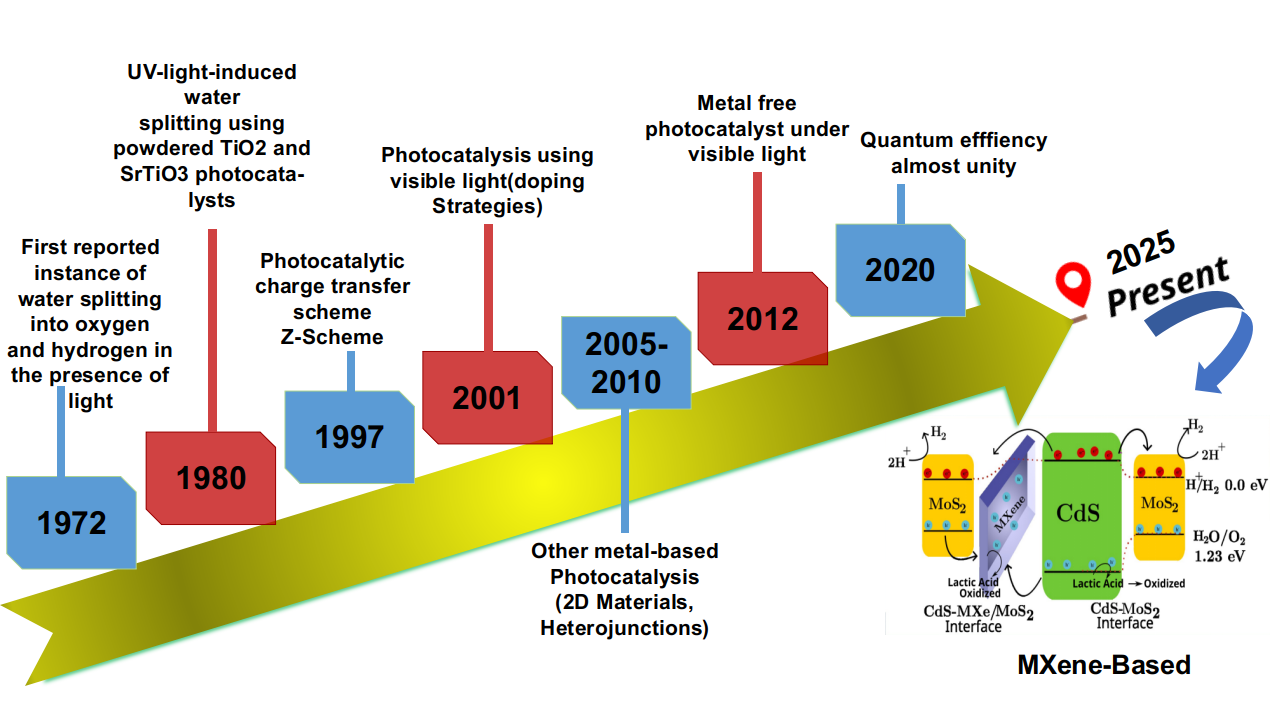}}
		\caption{Timeline showing the key milestones and material developments in the field of photocatalysis from early photochemical studies to modern advanced catalysts.}
		\label{milestone}
	\end{figure*}
	
	\subsection{Practical Considerations}
	Photocatalytic overall water splitting (OWS) is regarded as a promising strategy to fulfil future market demands for large-scale and cost-efficient hydrogen production.
	Photocatalysis enables efficient and effective chemical reactions at room temperature under sunlight irradiation and has attracted significant attention for its potential in developing ideal technologies that convert clean, safe, and abundant solar energy into electrical and/or chemical energy. A crucial pathway for achieving overall water splitting involves facilitating the four-electron transfer process~\cite{fujishima1972electrochemical}:
	\begin{equation}
		2\mathrm{H_2O} \rightarrow 2\mathrm{H_2} + \mathrm{O_2}
	\end{equation}
	and overcoming a substantial thermodynamic energy barrier of 1.8 eV, including the overpotential~\cite{xu2019visible,ma2014graphitic,yan2023electron,zhao2013nitrogen}. 
	\begin{figure}
		\centering
		\fbox{\includegraphics[width=\linewidth]{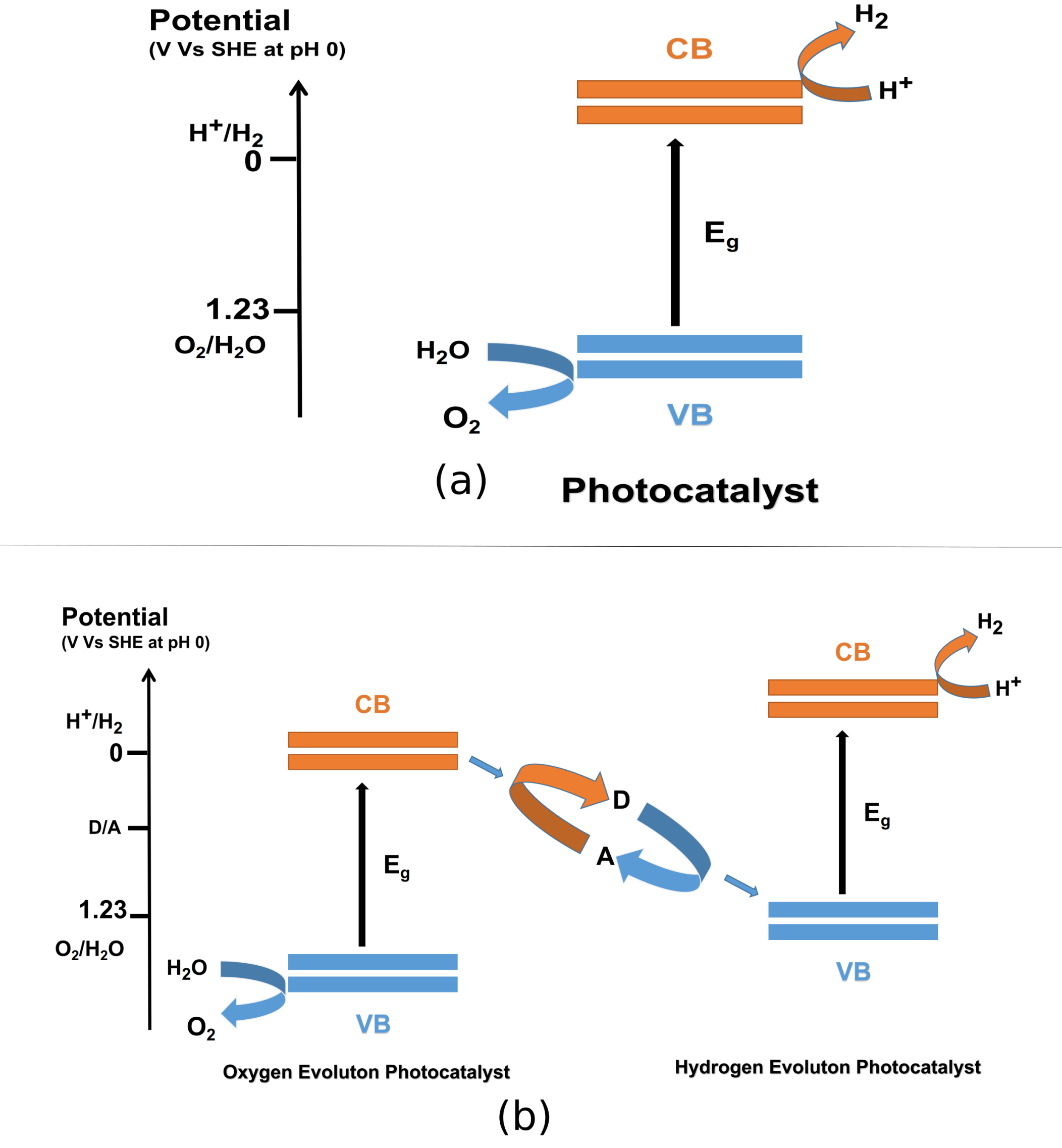}}
		\caption{(a) One-step photoexcitation system. (b) Two-step photoexcitation system}
		\label{model}
	\end{figure}
	
	Fig.~\ref{model} shows common schematic representations of energy band diagrams for water-splitting photocatalysts:  
	(a) Single-absorber photocatalyst system designed for overall water splitting.  
	(b) Dual-absorber photocatalyst system employing two semiconductors to enhance charge separation.  A \& D are redox couple(acceptor and donor species respectively), CB – conduction band,   
	$E_g$ – bandgap energy of the semiconductor, SHE – standard hydrogen electrode, VB – valence band. 
	The total minimum theoretical voltage required to split water under standard conditions is given by:
	
	\begin{equation}
		E^\circ_{\text{total}} = E^\circ_{\text{OER}} - E^\circ_{\text{SHE}} = 1.23\,\text{V} - 0\,\text{V} = 1.23\,\text{V}
	\end{equation}
	
	where $E^\circ_{\text{OER}} = 1.23\,\text{V}$ is the standard electrode potential for the oxygen evolution reaction (OER), and $E^\circ_{\text{SHE}} = 0\,\text{V}$ is the standard electrode potential of the Standard Hydrogen Electrode (SHE), which serves as the reference for the hydrogen evolution reaction (HER).
	
	However, due to kinetic barriers and overpotential, practical water splitting usually requires a higher potential (\textgreater 1.5 V)\cite{yan2023electron,bolton1985limiting,zeng2010recent}.
	
	The strategic design of advanced photocatalysts to modify the reaction kinetics and lower the thermodynamic energy barrier in \(\mathrm{H_2O}\) oxidation is crucial for efficient water splitting~\cite{wu2017control,srinivasan2016balanced,takanabe2017photocatalytic}.
	Tuning the electronic structures of active sites in metal-based catalysts has been 
	recognized as a promising strategy to enhance catalytic performance. 
	
	Wu \textit{et al.}~\cite{wu2017control} demonstrated a two-electron/two-step water splitting mechanism using transition-metal-doped structured catalysts. The incorporation of surface redox-active sites offers a direct and effective strategy to reduce the thermodynamic energy barrier associated with \(\mathrm{H_2O}\) oxidation.
	
	Photocatalytic water splitting involves three main steps: (1) light absorption, (2) charge carrier separation, and (3) surface redox reactions. The process is governed by the bandgap and band edge positions of the photocatalyst, which must straddle the hydrogen evolution reaction (HER) and oxygen evolution reaction (OER) potentials.
	
	One important efficiency metric is the external quantum efficiency (EQE), which is calculated using the following equation~\cite{nishioka2023photocatalytic}:
	\begin{equation}
		\text{EQE} (\%) = 2 \times \frac{N(\mathrm{H_2})}{N_{\text{photons}}} \times 100
	\end{equation}
	where \( N(\mathrm{H_2}) \) denotes the number of hydrogen molecules produced, and \( N_{\text{photons}} \) represents the number of incident photons reaching the surface of the reaction solution.
	
	The Solar-to-hydrogen (STH) efficiency is determined using the equation~\cite{yan2023electron}:
	\begin{align}
		\text{STH} (\%) = \frac{\text{Energy of generated } \mathrm{H_2}}{\text{Solar energy irradiating the reactor}} \times 100\% \notag \\
		= \frac{R_{\mathrm{H_2}} \times 4 \Delta G_r}{P_{\text{sun}} \times S} \times 100\%
	\end{align}
	
	Here, \( R_{\mathrm{H_2}} \) denotes the hydrogen production rate (mol/s), \( \Delta G_r \) represents the Gibbs free energy change for the formation of \(\mathrm{H_2}\), typically \( 237.13~\mathrm{kJ/mol} \), \( P_{\text{sun}} \) is the incident solar power density (\(\mathrm{W/m^2}\)), and \( S \) is the illuminated surface area (\(\mathrm{m^2}\)).
	
	Apparent Quantum Efficiency (AQE), also known as the apparent quantum yield (AQY)~\cite{yan2023electron}, is given by:
	\begin{align}
		\text{AQE} (\%) &= \frac{\text{Number of reacted electrons}}{\text{Number of incident photons}} \times 100\% \notag \\
		&= \frac{2 \times \text{Number of evolved } \mathrm{H_2} \text{ molecules}}{\text{Number of incident photons}} \times 100\% \notag \\
		&= \frac{(V \times N_A \times 2) \times (h \times c)}{(I \times A \times \lambda) \times t} \times 100\%
	\end{align}
	
	where \( \lambda \) is the wavelength of incident light (m), \( V \) is the hydrogen evolution activity (mol/h), \( N_A \) is Avogadro’s number, \( I \) is the light intensity (\(\mathrm{W/cm^2}\)), \( A \) is the illuminated area (\(\mathrm{cm^2}\)), \( t \) is the reaction time (s), \( h \) is Planck’s constant, and \( c \) is the speed of light.
	
	Hydrogen production through photocatalytic water splitting is a sustainable and efficient approach to addressing global energy demands. MXenes have recently gained attention due to their unique layered structures and favorable electronic properties. MXenes are a class of two-dimensional (2D) transition metal carbides, nitrides, and carbonitrides, with the general formula \( \mathrm{M_{n+1}X_nT_x} \),  
	where \( \mathrm{M} \) represents an early transition metal (e.g., Ti, V, Nb, Mo),  
	\( \mathrm{X} \) represents carbon (C) and/or nitrogen (N), and  
	\( \mathrm{T_x} \) represents surface functional groups such as –OH, –O, and –F.\\

	\textbf{Theoretical Work}: The selection of materials for photocatalysts, particularly for hydrogen production, must account for several critical theoretical factors~\cite{ran2017ti3c2}. Firstly, the photocatalyst should possess a suitable bandgap energy ($E_g$) in the range of 1.5 to 2.5~eV to enable efficient visible-light absorption. Moreover, for effective overall water splitting, the conduction band minimum (CBM) must be at a more negative potential than the hydrogen reduction potential (0~V vs. SHE), while the valence band maximum (VBM) should be more positive than the water oxidation potential (1.23~V vs. SHE).
	Efficient charge separation and transport are also essential; materials with high intrinsic conductivity, built-in electric fields, or heterojunction structures can markedly enhance the separation of photogenerated electron-hole pairs~\cite{xu2022design}. Additionally, stability under irradiation and in aqueous environments is crucial to mitigate photocorrosion and degradation during the reaction. Furthermore, the photocatalyst should have abundant and accessible surface active sites to facilitate the hydrogen evolution reaction (HER) and oxygen evolution reaction (OER). From a thermodynamic perspective, the Gibbs free energy of hydrogen adsorption ($\Delta G_{\text{H}^*}$) should be close to zero to ensure optimal catalytic activity without site blocking or weak interactions~\cite{jiao2015design}.
	
	\section{MXene-Based Photocatalysts: Structure and Properties}
	\begin{figure}
		\centering
		\fbox{\includegraphics[width=\linewidth]{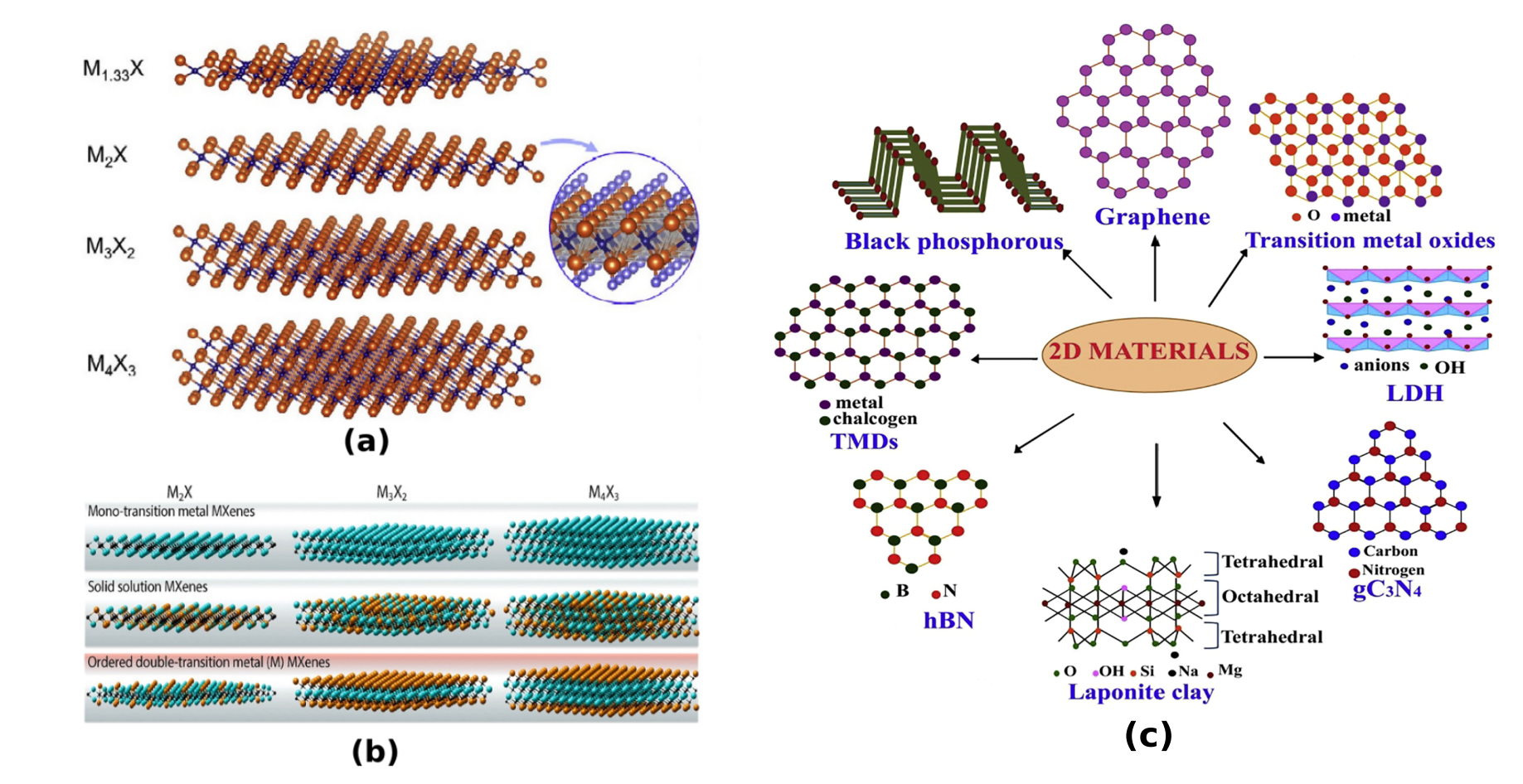}}
		\caption{(a) \& (b) Structure of different types of MXene. Reproduced with permission from Ref.~\cite{doi:10.1021/acsnano.9b06394,verger2019mxenes}, Copyright (2019) Elsevier. (c) Structure of other 2D materials, adapted with permission from Ref.~\cite{jayakumar20182d}, Copyright (2018) Elsevier.}
		\label{MXene}
	\end{figure}
	
	MXenes are synthesized through the selective etching of the A-layer from the MAX phase (a process extensively documented in various studies~\cite{BARSOUM2000201,alam2024advancements}), which follows the general formula M$_{n+1}$AX$_n$,  
	where A typically belongs to group 13 or 14 elements (e.g., Al, Si, Ga).  
	
	MXenes maintain a hexagonal close-packed (hcp) structure similar to their parent MAX phases.  
	The transition metal atoms (\( \mathrm{M} \)) form layered structures, with carbon or nitrogen atoms (\( \mathrm{X} \)) occupying octahedral interstitial sites.  
	Upon selective etching, the \( \mathrm{A} \)-layers are removed, resulting in a layered \( \mathrm{M\text{-}X} \) framework with surface terminations.
	
	The \( \mathrm{M\text{-}X} \) bond exhibits a hybrid character—covalent, metallic, and ionic—making it significantly stronger and structurally distinct from the weak van der Waals interactions found in traditional layered materials such as graphite and transition metal dichalcogenides (TMDs).  
	The interlayer spacing in MXenes can be tuned via ion intercalation, which critically influences their electronic and mechanical properties.
	
	The crystal structure of one of the most extensively studied MXenes, \( \mathrm{Ti_3C_2T_x} \), consists of:
	\begin{itemize}
		\item Titanium (\( \mathrm{Ti} \)) atoms arranged in a hexagonal lattice,
		\item Carbon (\( \mathrm{C} \)) atoms located in octahedral sites,
		\item Surface terminations (\( \mathrm{T_x} \)) such as –OH, –O, and –F that modulate the electronic and chemical characteristics.
	\end{itemize}
	Some of the different types of MXene materials as shown in Fig.\ref{MXene} (a) and (b) (where $n_1$:$n_2$ represents the ratio of transition metal (M) to carbon or nitrogen (X)) are listed below:
	\begin{itemize}
		\item \textbf{Single transition metal MXenes (2:1):} 
		\(\mathrm{Mo_2N}\)~\cite{urbankowski20172d}, 
		\((\mathrm{Ti}_{2-y}\mathrm{Nb}_y)\mathrm{C}\)~\cite{han2020tailoring}, 
		\(\mathrm{V_2C}\)~\cite{naguib2013new}, 
		\(\mathrm{Nb_2C}\)~\cite{naguib2013new}, 
		\(\mathrm{Mo_2C}\)~\cite{meshkian2015synthesis}, 
		\(\mathrm{Mo_{1.33}C}\)~\cite{persson2018tailoring}, 
		\(\mathrm{Ti_2N}\)~\cite{soundiraraju2017two}, 
		\((\mathrm{V}_{2-y}\mathrm{Nb}_y)\mathrm{C}\)~\cite{han2020tailoring}, 
		\(\mathrm{Mo_{1.33}Y_{0.67}C}\)~\cite{persson2018tailoring}, 
		\((\mathrm{Ti}_{2-y}\mathrm{V}_y)\mathrm{C}\)~\cite{han2020tailoring}, 
		\(\mathrm{W_{1.33}C}\)~\cite{meshkian2018w}, 
		\(\mathrm{Nb_{1.33}C}\)~\cite{halim2018synthesis}, and 
		\(\mathrm{Ti_2C}\)~\cite{naguib2012two}.
		
		\item \textbf{Single transition metal MXenes (3:2):} 
		\(\mathrm{Hf_3C_2}\)~\cite{zhou2017synthesis}, 
		\(\mathrm{Ti_3C_2}\)~\cite{naguib2023two}, 
		\(\mathrm{Zr_3C_2}\)~\cite{zhou2016two}, and 
		\(\mathrm{Ti_3CN}\)~\cite{naguib2012two}.
		
		\item \textbf{Single transition metal MXenes (4:3):} 
		\(\mathrm{Ta_4C_3}\)~\cite{naguib2012two}, 
		\(\mathrm{Nb_4C_3}\)~\cite{ghidiu2014synthesis}, 
		\(\mathrm{V_4C_3}\)~\cite{tran2018adding}, 
		\(\mathrm{Ti_4N_3}\)~\cite{urbankowski2016synthesis}, and 
		\((\mathrm{Mo},\mathrm{V})_4\mathrm{C}_3\)~\cite{pinto2020synthesis}.
		
		\item \textbf{Single transition metal MXenes (5:4):} 
		\(\mathrm{Mo_4VC_4}\)~\cite{deysher2019synthesis}.
		
		\item \textbf{Double transition metal MXenes (2:1:2):} 
		\(\mathrm{Mo_2TiC_2}\)~\cite{anasori2015two}, 
		\(\mathrm{Mo_2ScC_2}\)~\cite{meshkian2017theoretical}, and 
		\(\mathrm{Cr_2TiC_2}\)~\cite{anasori2015two}.
		
		\item \textbf{Double transition metal MXenes (2:2:3):} 
		\(\mathrm{Mo_2Ti_2C_3}\)~\cite{anasori2015two}.
	\end{itemize}
	
	MXenes possess exceptional properties, including high structural stability, remarkable electrochemical and optical characteristics, excellent thermal conductivity, superior electrical conductivity, and outstanding mechanical strength(extensively documented in various reviews~\cite{champagne2020physical,verger2019overview,verger2019mxenes,naguib2021ten}). These unique attributes stem from their diverse surface terminations and distinctive nanostructure, which combines covalent and metallic bonding. Furthermore, these properties can be tailored by altering surface functionalities, adjusting elemental composition, and optimizing the microstructure of MXenes~\cite{cao2022recent}.  
	
	MXenes surpass conventional 2D materials with abundant active sites, high Young’s modulus, excellent thermal and chemical stability, large interlayer spacing, and superior conductivity. These properties make them highly efficient photocatalysts, unlike traditional materials \cite{alhabeb2017guidelines,sinopoli2019electrocatalytic}. Surface functional groups further enhance their properties, enabling composite formation and bandgap tuning \cite{anasori20232d,ronchi2019synthesis}. Additionally, their high specific surface area (SSA) and mechanical stability contribute to their strong photocatalytic performance.
	The d-orbital electrons of transition metals influence adsorption energy, thereby  
	reducing the activation energy barrier and improving reaction efficiency~\cite{lykhach2016counting,campbell2012electronic} for photocatalytic applications.  
	\subsection{Electronic and Optical Properties}
	MXenes exhibit tunable band structures, high charge carrier mobility, and metallic-like conductivity, making them promising candidates for photocatalysis. One of their most notable properties is metallic conductivity, which varies depending on composition and surface functionalization. Some MXenes have also demonstrated semiconducting and superconducting behavior. Notably, \(\mathrm{Nb_2C}\) exhibits surface group-dependent superconductivity~\cite{kamysbayev2020covalent,anasori2016control}.
	
	The conductivity of MXene-based composites has been observed to surpass that of pristine MXenes~\cite{shen2018synthesis}. Berdiyorov et al.~\cite{berdiyorov2016optical} employed mathematical modeling to investigate how surface functionalization affects the optical properties of \(\mathrm{Ti_3C_2T_x}\) MXene. Their results showed that functionalization causes a twofold decrease in the static dielectric constant. Among various functional groups, the –O termination exhibits higher light absorption compared to the –F group and pristine MXene, primarily due to the impact of oxygen atoms on the total density of states near the Fermi level.
	
	Furthermore, surface functionalization enhances reflectance in the ultraviolet (UV) spectral range relative to unfunctionalized \(\mathrm{Ti_3C_2}\), while –O terminations increase both reflectance and absorption in the visible region. Oxygen and hydroxyl functionalizations further tune the optical response of MXenes. Owing to these exceptional optical properties, MXenes demonstrate remarkable photochemical conversion efficiencies~\cite{solangi2022mxene}.
	
	\subsection{Composition and Synthesis}
	\begin{figure*}
		\centering
		\fbox{\includegraphics[width=1.0\textwidth]{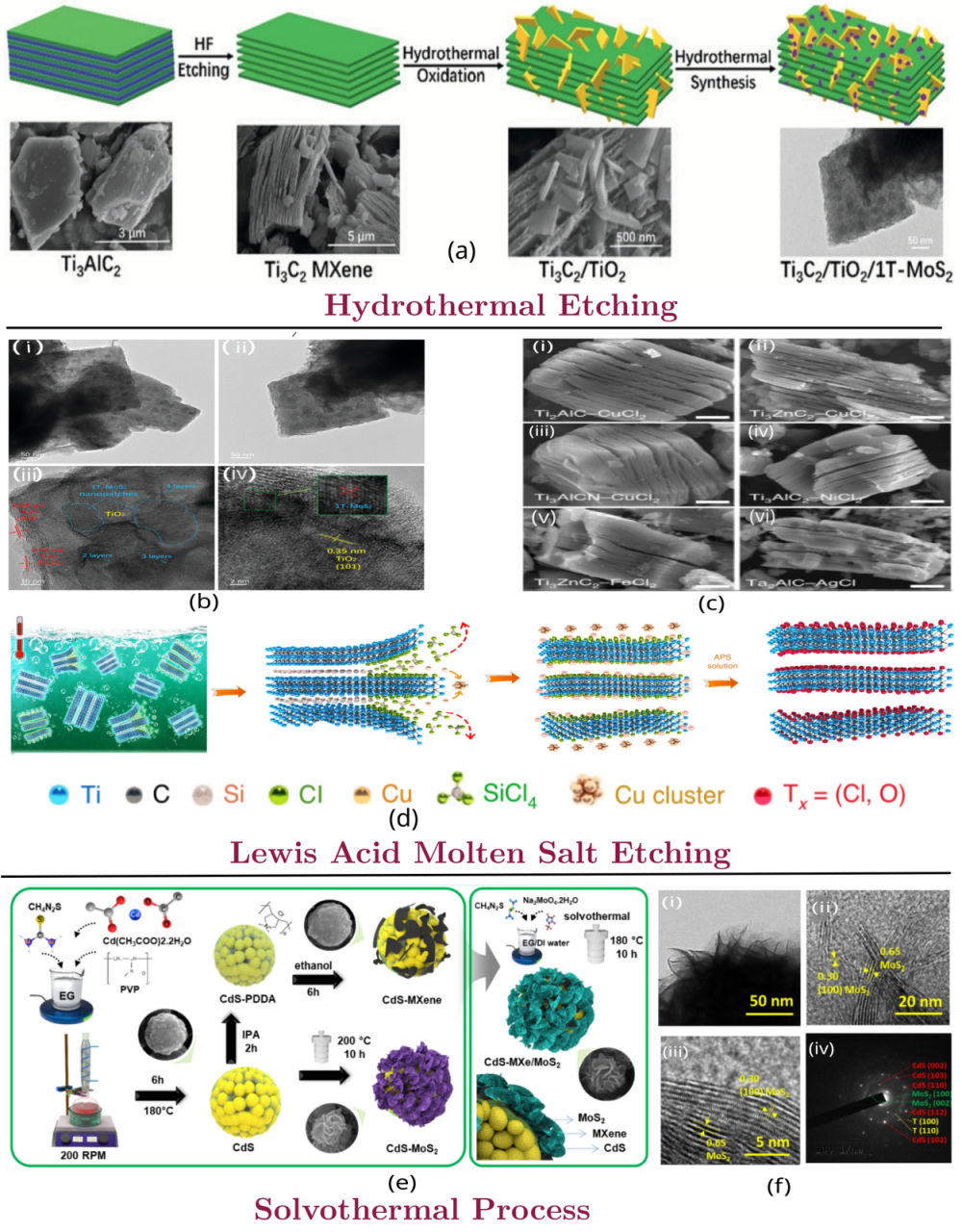}}
		\caption{Hydrothermal Etching (a)-(b) adapted with permission from Ref.~\cite{li20191t}, \copyright \,Copyright (2019) Royal Society of Chemistry.
			Lewis Acid Molten Salt Etching (c)-(d) Ref.~\cite{li2020general}, \copyright \,Copyright (2020) Nature Publishing Group UK London.
			Solvothermal Process (e)-(f) adapted with permission from Ref.~\cite{ranjith2024interfacial}, \copyright \,Copyright (2024) Springer.}
		\label{hydroLewis}
	\end{figure*}
	Synthesis methods of MXene includes chemical etching, in-situ growth, and exfoliation techniques.
	
\begin{figure}[htbp]
		\centering
		\fbox{\includegraphics[width=\linewidth]{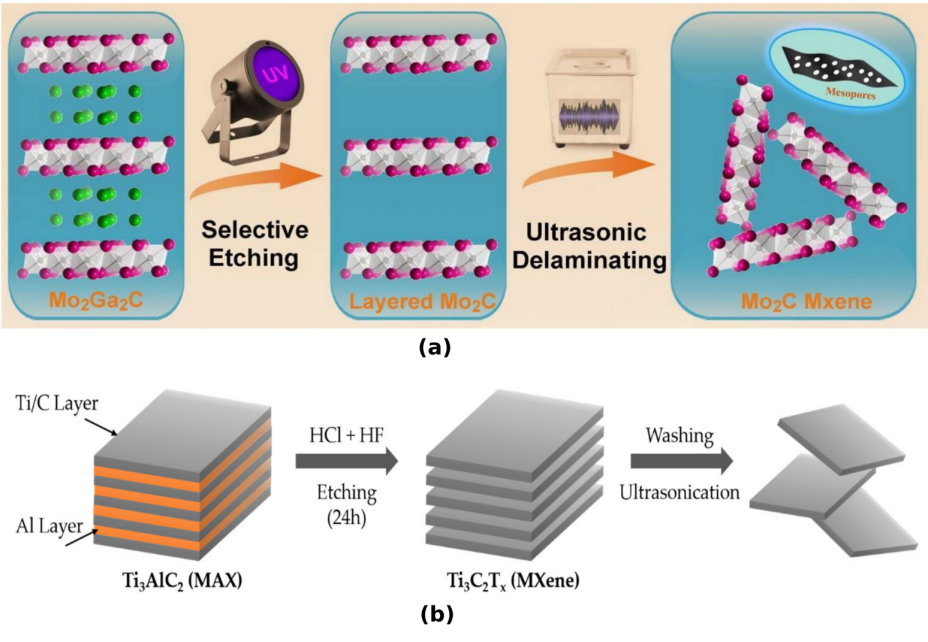}}
		\caption{(a) Schematic representation of the UV-assisted selective etching process for synthesizing two-dimensional (2D) mesoporous \(\mathrm{Mo_2C}\). (b) Schematic diagram illustrating the etching process of \(\mathrm{Ti_3AlC_2}\) into \(\mathrm{Ti_3C_2T_x}\). Reproduced with permission from Ref.~\cite{mei2020two} Copyright (2020) Elsevier and from Ref.~\cite{peng2021exploring} Copyright (2021) MDPI.}
		\label{etching}
	\end{figure}

	The first MXene, \(\mathrm{Ti_3C_2T_x}\), was synthesized by selectively etching the Al layer from the precursor MAX phase \(\mathrm{Ti_3AlC_2}\)~\cite{https://doi.org/10.1002/adma.201102306}, as shown in Fig.~\ref{etching}. Several etching methods using \(\mathrm{HCl/LiF}\) and HF salts, including \(\mathrm{LiF}\), \(\mathrm{NaF}\), \(\mathrm{KF}\), and \(\mathrm{NH_4F}\) in \(\mathrm{HCl}\), have been reported~\cite{natu20202d, husmann2020ionic}. However, chemical etching methods often introduce harmful impurities during the reaction. In contrast, physical preparation techniques mitigate these drawbacks, ensuring impurity-free synthesis.
	
	To overcome the limitations of fluoride-based etching, several fluoride-free methods have been developed. Yang et al.~\cite{yang2018fluoride} synthesized fluoride-free MXene titanium carbide (\(\mathrm{Ti_3C_2T_x}\), where \(T = \mathrm{O, OH}\)) using a binary aqueous system. The exfoliated sheets from this approach exhibited superior areal and volumetric capacitances in all-solid-state supercapacitors compared to those of LiF/HCl-etched MXenes. Several other fluoride-free synthesis techniques have been documented~\cite{li2018fluorine, sun2017electrochemical, pang2019universal, kumar2024fluorine}.
	
	Li \textit{et al.}~\cite{li20191t} reported the synthesis of a novel two-dimensional \(\mathrm{Ti_3C_2/TiO_2/1T\text{-}MoS_2}\) composite via a two-step hydrothermal method as shown in Fig.~\ref{hydroLewis}(a). Initially, \(\mathrm{TiO_2}\) nanosheets were grown \textit{in situ} on conductive \(\mathrm{Ti_3C_2}\) MXene, followed by uniform decoration with 1T-phase \(\mathrm{MoS_2}\) nanopatches. This strategy results in a unique composite structure featuring dual metallic co-catalysts—\(\mathrm{Ti_3C_2}\) and 1T-\(\mathrm{MoS_2}\)—with the 1T phase content reaching up to 84\%. The optimized composite with 15 wt\% \(\mathrm{MoS_2}\) loading exhibited a photocatalytic hydrogen evolution rate approximately 132, 11, and 1.5 times higher than that of pristine \(\mathrm{TiO_2}\) nanosheets, \(\mathrm{Ti_3C_2/TiO_2}\), and \(\mathrm{Ti_3C_2/TiO_2/2H\text{-}MoS_2}\) composites, respectively. The enhanced performance is attributed to the increased specific surface area and higher density of active sites due to 1T-\(\mathrm{MoS_2}\) nanopatches, along with improved electron transfer efficiency facilitated by the presence of both \(\mathrm{Ti_3C_2}\) MXene and metallic 1T-\(\mathrm{MoS_2}\).
	
	The morphology and microstructure of the \(\mathrm{Ti_3C_2/TiO_2/1T\text{-}MoS_2}\) composites were investigated using transmission electron microscopy (TEM), as illustrated in Fig.~\ref{hydroLewis}(b)-(i) and (b)-(ii). High-resolution TEM (HRTEM) images in Fig.~\ref{hydroLewis}(b)-(iii) and (b)-(iv) reveal detailed structural characteristics at the interface between 1T-\(\mathrm{MoS_2}\) and the \(\mathrm{TiO_2}\)–\(\mathrm{Ti_3C_2}\) hybrid. Moreover, a lattice spacing of 0.35~nm corresponds to the (101) plane of anatase \(\mathrm{TiO_2}\), as shown in Fig.~\ref{hydroLewis}(b)-(iv). The \(\mathrm{MoS_2}\) nanopatches consist of approximately 2 to 5 layers, as depicted in Fig.~\ref{hydroLewis}(b)-(iii).
	
	Li \textit{et al.}~\cite{li2019element} recently reported the synthesis of the \(\mathrm{Ti_3ZnC_2}\) MAX phase through a replacement reaction, where \(\mathrm{Ti_3AlC_2}\) was reacted with \(\mathrm{ZnCl_2}\) in a Lewis acidic molten salt medium at 550~$^\circ$C. By increasing the MAX:\(\mathrm{ZnCl_2}\) ratio, \(\mathrm{Ti_3ZnC_2}\) can be further converted into \(\mathrm{Ti_3C_2Cl_2}\) MXene.
	
	Fig.~\ref{hydroLewis}(d) presents a schematic illustration of \(\mathrm{Ti_3C_2T_x}\) MXene synthesis. The \(\mathrm{Ti_3SiC_2}\) MAX phase is immersed in \(\mathrm{CuCl_2}\) Lewis molten salt at 750~$^\circ$C, where a reaction between \(\mathrm{Ti_3SiC_2}\) and \(\mathrm{CuCl_2}\) results in the formation of \(\mathrm{Ti_3C_2T_x}\) MXene. The resulting material, referred to as MS-\(\mathrm{Ti_3C_2T_x}\) MXene, is obtained following a post-treatment step involving washing with ammonium persulfate (APS) solution~\cite{li2020general}. Scanning electron microscopy (SEM) images in Fig.~\ref{hydroLewis}(d) reveal the distinct accordion-like structures typical of MXenes, produced by etching various MAX phases with different Lewis acidic chloride salts. Specifically, \(\mathrm{Ti_2AlC}\) was etched using \(\mathrm{CuCl_2}\) (Fig.~\ref{hydroLewis}(c)-i), \(\mathrm{Ti_3ZnC_2}\) with \(\mathrm{CuCl_2}\) (c)-ii, \(\mathrm{Ti_3AlCN}\) with \(\mathrm{CuCl_2}\) (c)-iii, \(\mathrm{Ti_3AlC_2}\) with \(\mathrm{NiCl_2}\) (c)-iv, \(\mathrm{Ti_3ZnC_2}\) with \(\mathrm{FeCl_2}\) (c)-v, and \(\mathrm{Ta_2AlC}\) with \(\mathrm{AgCl}\) (c)-vi. Each image includes a scale bar of 2~\(\mu\)m.
	
	Hydrothermal and Lewis acid etching methods, as shown in Fig.~\ref{hydroLewis}(a)-(b), enhance safety, efficiency, and environmental sustainability. These provide more efficient, reliable, and time-saving alternatives to HF etching~\cite{li2019element}. The Lewis acid method enables the synthesis of diverse MXenes while incorporating –Cl and –Br functional groups. Kulkarni et al.~\cite{kulkarni2024hydrothermal} synthesized MXene via hydrothermal etching of the MAX phase (\(\mathrm{Ti_3AlC_2}\)) using relatively non-toxic alkaline solutions such as potassium hydroxide. Additionally, chemical vapor deposition (CVD) has emerged as a viable synthesis approach for producing ultra-thin MXene layers~\cite{hong2020chemical, xu2015large, geng2017direct, liu2016unique, salim2019introduction}.
	
	The solvothermal process, as shown in Fig.~\ref{hydroLewis}(e), is another safe and efficient method for synthesizing ternary nanospheres, as reported by Ranjith \textit{et al.}~\cite{ranjith2024interfacial}. This study presents a \(\mathrm{CdS}\)–MXene/\(\mathrm{MoS_2}\) ternary core–shell heterostructure, where MXene layers enhance hole capture and \(\mathrm{MoS_2}\) nanosheets act as co-catalysts to facilitate efficient charge separation and hydrogen evolution. The composite achieved a high \(\mathrm{H_2}\) production rate of 38,500~\(\mu\)mol~g\(^{-1}\)~h\(^{-1}\) under visible light, outperforming binary and single-component systems, with an apparent quantum efficiency (AQE) of 34.6\% at 420~nm.
	Field emission transmission electron microscopy (FETEM) and high-resolution transmission electron microscopy (HRTEM) images, shown in Fig.~\ref{hydroLewis}(e)-(i \& ii) for \(\mathrm{CdS}\)–MXene and (e)-(iii) for \(\mathrm{CdS}\)–MXene/\(\mathrm{MoS_2}\), provide detailed insights into the structural and interfacial characteristics of the heterostructure. Fig.~\ref{hydroLewis}(f)-(iv) displays the selected area electron diffraction (SAED) pattern of\(\mathrm{CdS}\)–MXene/\(\mathrm{MoS_2}\).
	These emerging synthesis strategies present highly promising and efficient avenues for producing MXenes with optimized properties for advanced photocatalytic applications~\cite{ran2017ti3c2,liu2020one,qin2022hydrothermal,siddique2023fluorine,huang2023recent}. It is crucial to recognize that the properties and applications of MXenes are significantly influenced by the chosen synthesis technique. The development of effective surface functional groups plays a vital role in determining the performance and suitability of MXene materials for various applications~\cite{hui2020interface,li2022mxene,murali2022review}.
	
	Table~\ref{all} presents the influence of various synthesis routes on MXene characteristics and their respective applications. Recent advancements have introduced several new synthesis methods for MXenes, which have been extensively summarized in various studies~\cite{kumar2024fluorine,wei2021advances,naguib2021ten,jayakumar20242d,rostami2024review}. These methods further enhance the structural control, stability, and functionalization of MXenes, expanding their potential applications in energy storage, catalysis, and optoelectronics.
	
	\section{Recent Advances in MXene-Based Photocatalysts}
	\subsection{MXene/Semiconductor Composites/Metal Co-Catalysts}
	
	\begin{figure*}
		\centering
		\fbox{\includegraphics[width=\linewidth]{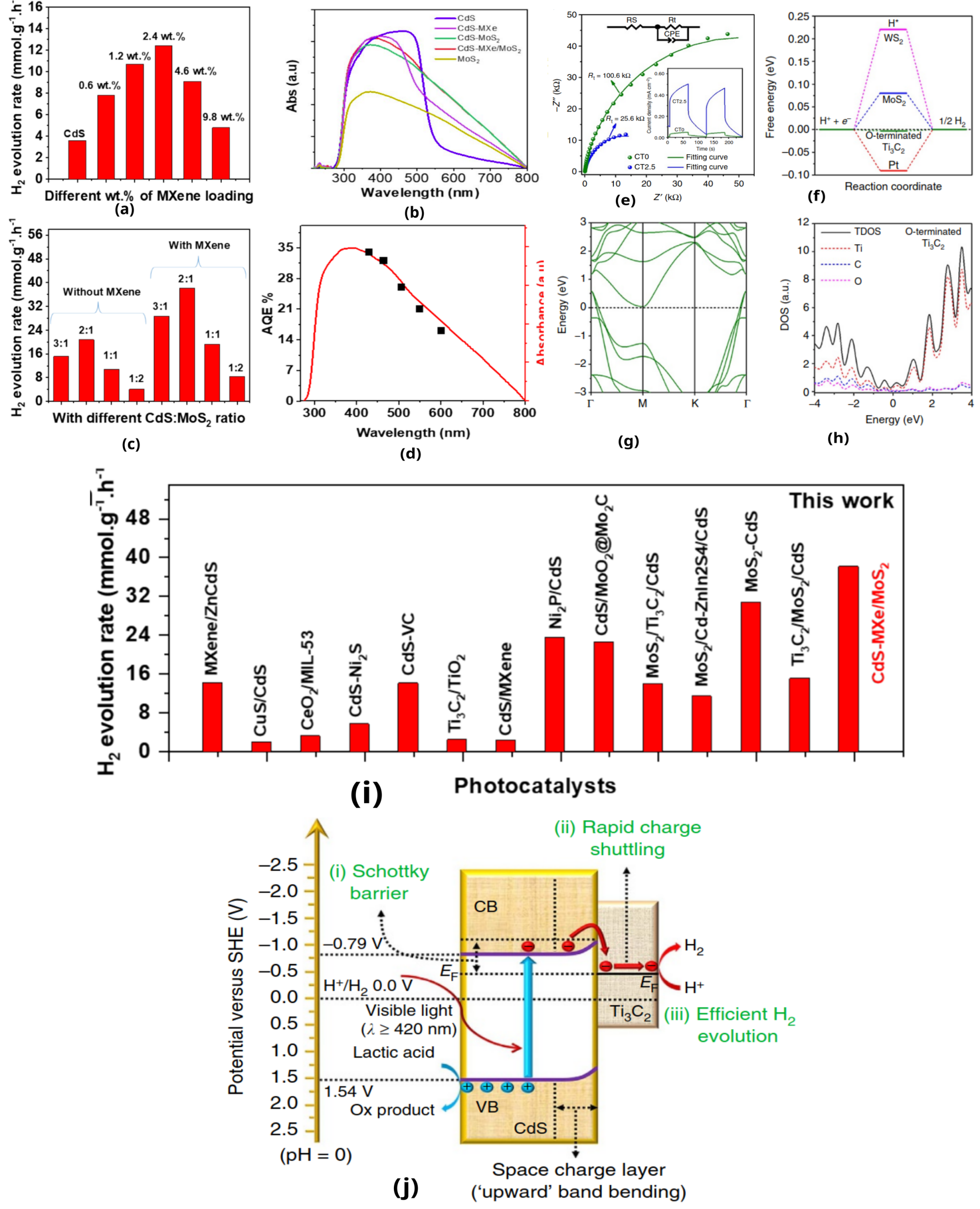}}
		\caption{(a) \(\mathrm{H_2}\) production with different wt.\% of MXene loading, (b) UV-DRS spectra of \(\mathrm{CdS}\), \(\mathrm{CdS}\)–MXene, and \(\mathrm{CdS}\)–MXene/\(\mathrm{MoS_2}\) heterostructures, (c) \(\mathrm{H_2}\) production with different \(\mathrm{CdS}\):\(\mathrm{MoS_2}\) ratios, (d) apparent quantum efficiency (AQE) at different wavelengths, (e) electrochemical impedance spectroscopy (EIS) Nyquist plots, (f) free energy profile of the hydrogen evolution reaction (HER) at equilibrium potential (\(U=0\,\mathrm{V}\)), (g) and (h) band structure and density of states (DOS) of oxygen-terminated \(\mathrm{Ti_3C_2}\). Adapted with permission from Ref.~\cite{ranjith2024interfacial}, \copyright\, Copyright (2024) Springer. Adapted with permission from Ref.~\cite{ran2017ti3c2}, \copyright\, Copyright (2017) Nature Publishing Group UK London.}
		\label{All3Table16and12}
	\end{figure*}
	The application of MXene-based photocatalysts for the degradation of toxic pollutants, \(\mathrm{CO}_2\) reduction, and other environmental remediation processes has been extensively reviewed in the literature~\cite{solangi2023mxene,jin2023recent,aldeen2023environmental,wang2025photocatalytic,vasseghian2022comprehensive}. However, this paper primarily focuses on their role in water splitting and hydrogen production.
	
	MXenes, owing to their high specific surface area~\cite{hou2020ti}, metallic conductivity~\cite{zhang2023scalable,liu2020ultrastrong,zhang2021flexible}, and hydrophilicity~\cite{han2017preparation}, exhibit significant potential as cocatalysts for \(\mathrm{TiO}_2\)-based photocatalysis~\cite{zhong2021two}. Their excellent conductivity enables them to function as electron reservoirs, leading to the formation of a Schottky barrier at the \(\mathrm{TiO}_2\) interface, which enhances charge separation and transfer~\cite{wang2020single}. Among MXenes, \(\mathrm{Ti}_3\mathrm{C}_2\) is the most extensively studied due to its superior structural stability and electrical conductivity.
	
	In Fig.~\ref{All3Table16and12}(a), the photocatalytic performance of pristine \(\mathrm{CdS}\) with varying MXene loading densities was investigated~\cite{ranjith2024interfacial}. The \(\mathrm{H}_2\) evolution rate of bare \(\mathrm{CdS}\) was relatively low, at 3,600~\(\mathrm{\mu mol\,g^{-1}\,h^{-1}}\), primarily due to particle aggregation and the rapid recombination of photogenerated electron–hole pairs. By increasing the MXene content from 0.6 to 2.4~wt\%, the \(\mathrm{H}_2\) generation rate significantly improved, rising from 7,800 to 12,400~\(\mathrm{\mu mol\,g^{-1}\,h^{-1}}\). These results confirm the enhancement effect of MXene, which promotes charge carrier migration and reduces recombination losses on the \(\mathrm{CdS}\) surface.
	
	As shown in Fig.~\ref{All3Table16and12}(b), \(\mathrm{CdS}\) nanospheres exhibited strong visible light absorption at 549~nm, which was extended further into the visible region by tagging MXene. Increasing MXene loading (0.6–4.8~wt\%) enhanced light absorption in the 550–800~nm range, while the addition of \(\mathrm{MoS}_2\) further promoted light absorption through the synergistic effect of the heterostructure.
	
	Moreover, the ternary \(\mathrm{CdS}\)–MXene/\(\mathrm{MoS}_2\) heterostructure demonstrated a significantly enhanced photocatalytic \(\mathrm{H}_2\) evolution rate, being 3.1 and 1.9 times higher than those of the \(\mathrm{CdS}\)–MXene and \(\mathrm{CdS}/\mathrm{MoS}_2\) binary heterostructures, respectively. The role of MXene as an interlayer was systematically investigated by adjusting the \(\mathrm{MoS}_2\) growth density on the \(\mathrm{CdS}\) surface, as shown in Fig.~\ref{All3Table16and12}(c). It was found that excessive \(\mathrm{MoS}_2\) loading negatively impacted photocatalytic activity due to a reduction in active sites, increased light shielding, greater charge recombination, and deterioration of the core–shell morphology. An optimal \(\mathrm{CdS}:\mathrm{MoS}_2\) mass ratio of 2:1, combined with a 2.4~wt\% MXene loading, created a synergistic effect that significantly improved the photocatalytic performance.
	
	Fig.~\ref{All3Table16and12}(b) presents the photocatalytic \(\mathrm{H}_2\) production rates of \(\mathrm{CdS}\), \(\mathrm{CdS}\)–MXene\(_{2.4}\), and \(\mathrm{CdS}\)–MXene\(_{2.4}/\mathrm{MoS}_2\) heterostructures, demonstrating that MXene integration significantly improved the \(\mathrm{H}_2\) evolution rate, while subsequent \(\mathrm{MoS}_2\) decoration further enhanced the performance. As shown in Fig.~\ref{All3Table16and12}(a), \(\mathrm{MoS}_2\) decoration alone increased the \(\mathrm{H}_2\) production rate of \(\mathrm{CdS}\)–\(\mathrm{MoS}_2\) to 20,800~\(\mathrm{\mu mol\,g^{-1}\,h^{-1}}\). Owing to the synergistic effect of the interlayered MXene, the hierarchical \(\mathrm{CdS}\)–MXene/\(\mathrm{MoS}_2\) heterostructure achieved an outstanding \(\mathrm{H}_2\) production rate of 38,500~\(\mathrm{\mu mol\,g^{-1}\,h^{-1}}\), representing a 10.7-fold improvement compared to pristine \(\mathrm{CdS}\).
	
	The apparent quantum yield (AQY) of the \(\mathrm{CdS}\)–MXene\(_{2.4}/\mathrm{MoS}_2\) catalyst achieved values of 34.6\%, 32.0\%, 26.3\%, 21.4\%, and 16.2\% at wavelengths of 420, 460, 510, 560, and 620~nm, respectively, as depicted in Fig.~\ref{All3Table16and12}(d), exceeding the performance of pristine \(\mathrm{CdS}\) across the visible spectrum.
	
	Compared to other reported catalysts, the \(\mathrm{CdS}\)–MXene/\(\mathrm{MoS}_2\) heterostructure exhibited superior catalytic activity and stability for photocatalytic \(\mathrm{H}_2\) production, as shown in Fig.~\ref{All3Table16and12}(i).
	
	\begin{figure*}
		\centering    \fbox{\includegraphics[width=1.0\textwidth]{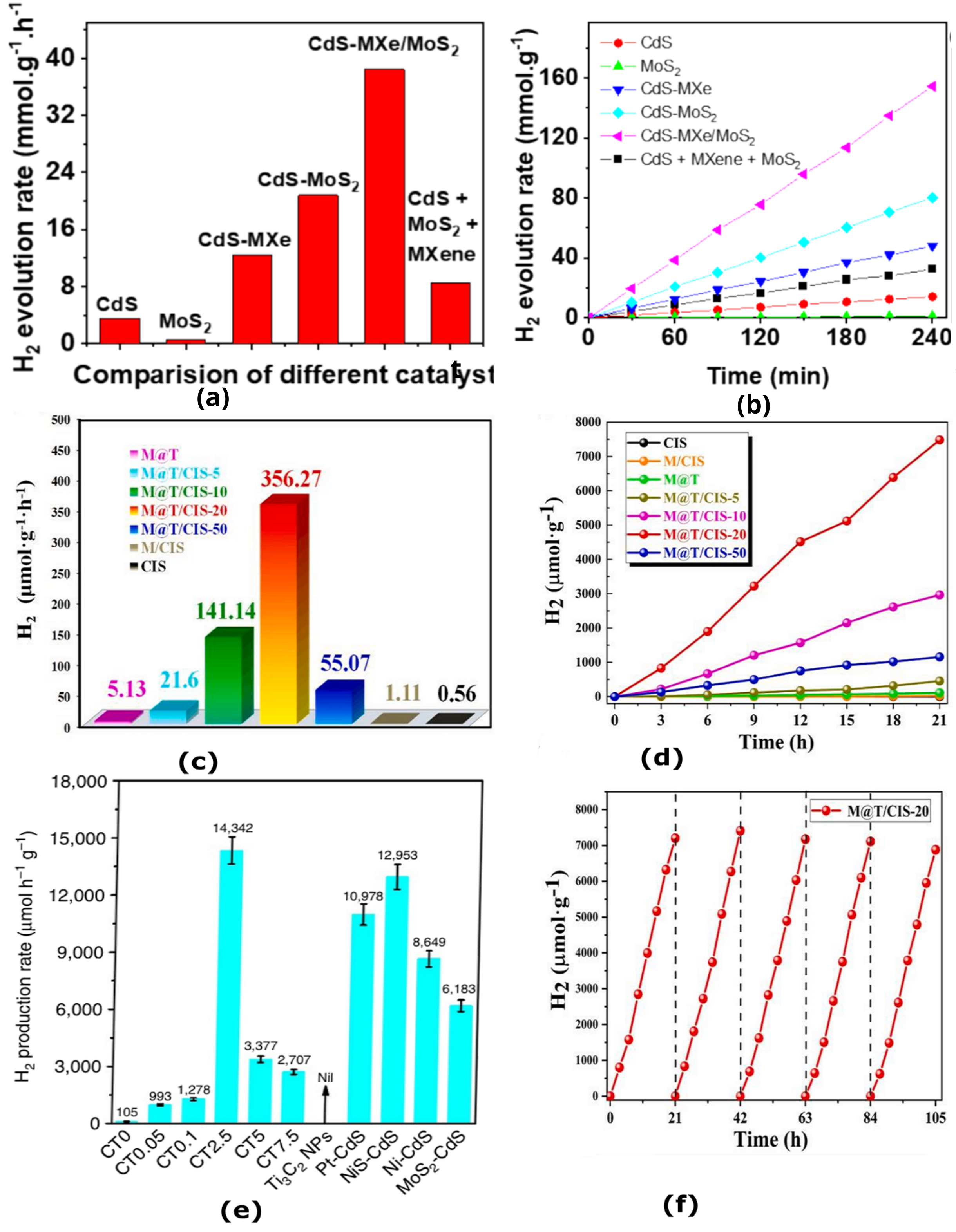}}
		\caption{(a) \& (b) Adapted with permission from Ref.~\cite{ranjith2024interfacial} Copyright (2024)  Springer. (c),(d) \&(f)Ref.~\cite{yang2022rationally} Copyright (2022)  Elsevier.
			(e) Ref.~\cite{ran2017ti3c2} Copyright (2017)  Nature Publishing Group UK London.}
		\label{all}
	\end{figure*}

\begin{figure}[H]
		\centering
		\fbox{\includegraphics[width=\linewidth]{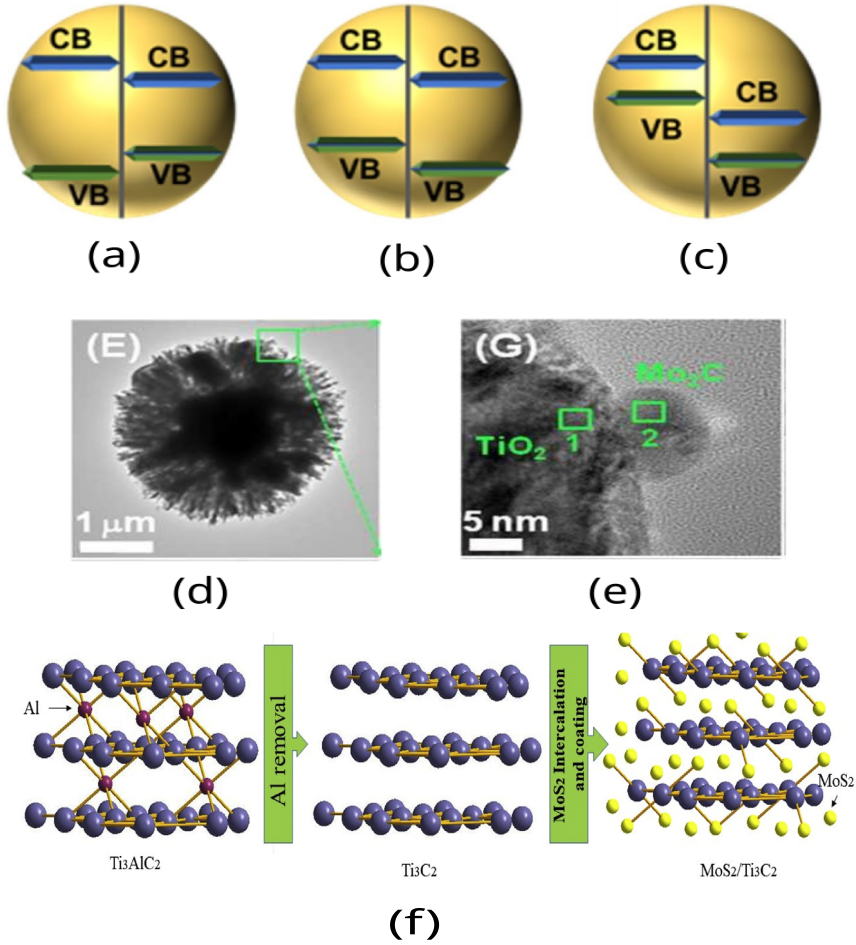}}
		\caption{(a)-(c) Various types of heterostructure. (d)-(e) Transmission electron microscopy (TEM) image of \(\mathrm{Mo}_2\mathrm{C}@\mathrm{TiO}_2\) and high-resolution TEM (HRTEM) images of 1 wt\% \(\mathrm{Mo}_2\mathrm{C}/\mathrm{TiO}_2\). 
			Adapted with permission from Ref.~\cite{yue2017novel}, Copyright (2017) Royal Society of Chemistry.
			(f) Schematic representation of the synthesis process of the \(\mathrm{MoS}_2/\mathrm{Ti}_3\mathrm{C}_2\) heterostructure. Adapted with permission from Ref.~\cite{zhang2020mos2}, Copyright (2020) Elsevier.}
		\label{heterostructure}
	\end{figure}

	Heterojunctions of MXenes with semiconductors such as \(\mathrm{TiO}_2\), \(\mathrm{g}\text{-}\mathrm{C}_3\mathrm{N}_4\), and \(\mathrm{CdS}\) is an effective approach to enhance charge separation and broaden light absorption as shown in Fig.~\ref{heterostructure}. A heterojunction is a junction formed between two different semiconductor materials with suitable band alignment. This approach enhances charge separation and transfer, reducing electron-hole recombination and improving photocatalytic efficiency. Fig.~\ref{heterostructure} (a)-(c) are schematic of different types of heterostructures. The SEM images in (d) \& (e) clearly show that \(\mathrm{Mo}_2\mathrm{C}\) nanoparticles (NPs) with diameters less than 10~nm were uniformly grown on the \(\mathrm{TiO}_2\) surface and (f) illustrates the synthesis process of heterostructures.
	
	Yang \textit{et al}.~\cite{yang2022rationally} reported the synthesis of a ternary \(\mathrm{Ti}_3\mathrm{C}_2\) MXene/\(\mathrm{TiO}_2\)/\(\mathrm{CuInS}_2\) (M@T/CIS) heterojunction photocatalyst via a two-step hydrothermal in-situ growth method. The optimized M@T/CIS composite achieved a high \(\mathrm{H}_2\) production rate of 356.27~\(\mu\mathrm{mol\,g^{-1}\,h^{-1}}\), as illustrated in Fig.~\ref{all}(c) \& (d). Notably, the M@T/CIS-20 sample maintained a stable \(\mathrm{H}_2\) evolution rate over six consecutive photocatalytic cycles (105~h), indicating excellent reusability and long-term durability. This superior performance is attributed to enhanced light absorption, effective charge separation facilitated by S-scheme and Schottky junctions, and the presence of abundant active sites from \(\mathrm{Ti}_3\mathrm{C}_2\) MXene.
	
	Juhui Zhang \textit{et al}.~\cite{zhang2020mos2} reported that the \(\mathrm{MoS}_2/\mathrm{Ti}_3\mathrm{C}_2\) heterostructure, with a 30\% \(\mathrm{Ti}_3\mathrm{C}_2\) loading, attained a maximum hydrogen production rate of 6144.7~\(\mathrm{mmol\,g^{-1}\,h^{-1}}\), exceeding that of pure \(\mathrm{MoS}_2\) by a factor of 2.3. Integration of MXenes with noble metals (Pt, Au) improves the hydrogen evolution reaction by reducing charge recombination.
	
	Kuang \textit{et al}.~\cite{kuang2016embedding} demonstrated that \(\mathrm{Ti}_3\mathrm{C}_2\) MXene, when used as a cocatalyst, plays a crucial role in enhancing photocatalytic performance. Its excellent electrical conductivity facilitates rapid electron transfer, while its high chemical stability ensures durability under photocatalytic conditions. Additionally, the large specific surface area of \(\mathrm{Ti}_3\mathrm{C}_2\) provides abundant active sites, further improving catalytic efficiency. By leveraging the Schottky junction mechanism, the \(\mathrm{CdS}/\mathrm{Ti}_3\mathrm{C}_2\) system effectively accelerates charge separation, thereby significantly enhancing its effectiveness in applications such as water splitting\cite{sun2021fabrication}.
	
	Employing co-catalysts is a well-established method to enhance the efficiency of photocatalysts in water splitting. M. Shao \textit{et al}.~\cite{shao2019carbonized} demonstrated that carbonized \(\mathrm{MoS}_2\) (\(\mathrm{MoS}_2/\mathrm{Mo}_2\mathrm{C}\)) acts as an exceptionally active co-catalyst for solar-driven hydrogen production. The \(\mathrm{MoS}_2/\mathrm{Mo}_2\mathrm{C}\)-modified \(\mathrm{CdS}\) catalyst exhibits an exceptionally high photocatalytic hydrogen evolution rate of 34,000~\(\mu\mathrm{mol\,g^{-1}h^{-1}}\), which is approximately 112 times greater than that of pure \(\mathrm{CdS}\). Additionally, it demonstrates an outstanding apparent quantum efficiency (AQE) of 41.4\% at 420~nm.
	
	According to Ran \textit{et al}.~\cite{ran2014earth}, a highly active co-catalyst can not only efficiently extract photoinduced electrons from the photocatalyst to its surface but also effectively catalyze \(\mathrm{H}_2\) evolution using these electrons. The Gibbs free energy of intermediate adsorbed state, \(|\Delta G_{\mathrm{H}^*}|\), is widely recognized as a fundamental parameter for assessing HER catalytic performance, with an ideal value approaching zero.\cite{jiao2015design} Notably, platinum (Pt), a highly efficient HER catalyst, demonstrates a near-optimal value of \(\Delta G_{\mathrm{H}^*} \approx - 0.09\,\mathrm{eV}\) \cite{hinnemann2005biomimetic}.
	
	Ran \textit{et al}.~\cite{ran2017ti3c2} successfully conducted first-principles theoretical investigations and reported that oxygen-functionalized \(\mathrm{Ti}_3\mathrm{C}_2\) satisfies the criteria for an excellent co-catalyst, as shown in Fig.~\ref{All3Table16and12}(f), performing even better than the commonly used Pt. Furthermore, Fig.~\ref{All3Table16and12}(g) and (h) illustrate its favorable band structure and high conductivity. The Fermi levels (\(E_{\mathrm{F}}\)) of \(\mathrm{Ti}_3\mathrm{C}_2\), O-terminated \(\mathrm{Ti}_3\mathrm{C}_2\), and F-terminated \(\mathrm{Ti}_3\mathrm{C}_2\) were determined to be \(-0.05\,\mathrm{V}\), \(1.88\,\mathrm{V}\), and \(0.15\,\mathrm{V}\) versus the standard hydrogen electrode (SHE), respectively. Notably, O-terminated \(\mathrm{Ti}_3\mathrm{C}_2\) exhibits the most positive \(E_{\mathrm{F}}\), suggesting an enhanced capability to accept photo-induced electrons from semiconductor photocatalysts. Experimentally, this material demonstrated outstanding visible-light photocatalytic performance, achieving an \(\mathrm{H}_2\) evolution rate of 14,342~\(\mu\mathrm{mol\,g^{-1}\,h^{-1}}\) as shown in Fig.~\ref{all}(e) and an apparent quantum efficiency (AQE) of 40.1\% at 420~nm, positioning it among the most efficient noble-metal-free metal sulfide photocatalysts. As illustrated in Fig.~\ref{All3Table16and12}(e) through the EIS Nyquist plots, the \(\mathrm{CdS}/\mathrm{Ti}_3\mathrm{C}_2\) composite with 2.5~wt\% \(\mathrm{Ti}_3\mathrm{C}_2\) loading (CT2.5) demonstrates a much smaller semicircle and significantly reduced interfacial charge-transfer resistance compared to bare \(\mathrm{CdS}\) (CT0) under visible-light irradiation in potassium phosphate buffer solution (pH = 7), signifying improved interfacial charge-carrier transfer at the \(\mathrm{CdS}/\mathrm{Ti}_3\mathrm{C}_2\) interface.
	
	Fig.~\ref{All3Table16and12} (j) shows a schematic illustration of the charge separation and transfer mechanism in the \(\mathrm{CdS}/\mathrm{Ti}_3\mathrm{C}_2\) system under visible-light irradiation, where red and blue spheres represent photo-induced electrons and holes, respectively.
	
	\begin{table}[htbp]
		\caption{Summary of photocatalytic hydrogen evolution performance of various MXene-based photocatalysts using different synthesis methods, characterization, and efficiency.}
		\label{photocatalyst_summary}
		\begin{tabular}{|p{2.50cm}|p{1.5cm}|p{2.25cm}|p{1.8cm}|p{2.0cm}|p{0.8cm}|p{2.25cm}|p{0.35cm}|}
			\hline
			\textbf{MXene-based Photocatalyst} & \textbf{Synthesis Method} & \textbf{Characterization Techniques} & \textbf{Light Source} & \textbf{\(\mathrm{H}_2\) Evolution Rate} & \textbf{AQE (\%)} & \textbf{Remarks} & \textbf{Ref} \\
			\hline
			MXene/\(\mathrm{MoS}_2\)@CdS & Solvothermal & XRD, TEM, HRTEM, SEM, XPS, UV-Vis & Xenon lamp & 38,500 \(\mu\mathrm{mol\,g^{-1}\,h^{-1}}\) & 34.6 & Ternary heterostructure & \cite{ranjith2024interfacial} \\
			\hline
			\(\mathrm{Ti}_3\mathrm{C}_2/\mathrm{TiO}_2/\mathrm{CuInS}_2\) & Hydrothermal & XRD, TEM, HRTEM, SEM, XPS & Visible (Labsolar-6A) & 356.27 \(\mu\mathrm{mol\,g^{-1}\,h^{-1}}\) & 1.9 & S-scheme and Schottky junctions & \cite{yang2022rationally} \\
			\hline
			MXene–\(\mathrm{CdS}/\mathrm{WO}_3\) & Solvothermal & XRD, TEM, HRTEM, SEM, XPS, PL-SPS, UPS & Sunlight (Xe lamp) & 27,500 \(\mu\mathrm{mol\,g^{-1}\,h^{-1}}\) & 12.0 & S-scheme heterojunction & \cite{bai2022integration} \\
			\hline
			\(\mathrm{Ti}_3\mathrm{C}_2/\mathrm{CdS}/\mathrm{ZnS}\) & Hydrothermal & HAADF, TEM, HRTEM, PDOS, TDOS, EDX, SEM & 300 W Xenon arc lamp & 14,342 \(\mu\mathrm{mol\,g^{-1}\,h^{-1}}\) & 40.1 & High visible-light photocatalytic activity & \cite{ran2017ti3c2} \\
			\hline
			\(\mathrm{MoS}_2/\mathrm{Ti}_3\mathrm{C}_2\) & HF etching \& Hydrothermal & XRD, TEM, HRTEM, SEM, XPS & Visible & 6144.7 \(\mu\mathrm{mol\,g^{-1}\,h^{-1}}\) & – & Enhanced activation sites & \cite{zhang2020mos2} \\
			\hline
			\(\mathrm{Ti}_3\mathrm{C}_2/\mathrm{CdS}\) & Hydrothermal & XRD, SEM, TEM, HRTEM, XPS & Visible & 63.53 \(\mu\mathrm{mol\,h^{-1}}\) & 2.28 & Co-catalyst strategy & \cite{sun2021fabrication} \\
			\hline
			\(\mathrm{SnS}_2/\mathrm{Ti}_3\mathrm{C}_2/\mathrm{TiO}_2\) (STT) & Hydrothermal & XRD, SEM, XPS, UV-Vis & Visible (300 W Xe arc lamp) & 10,505.8 \(\mu\mathrm{mol\,g^{-1}\,h^{-1}}\) & – & Defect-rich \(\mathrm{SnS}_2\) and \(\mathrm{Ti}_3\mathrm{C}_2\) MXene as cocatalysts & \cite{varadarajan2024enhanced} \\
			\hline
			\(\mathrm{MoS}_2\) (\(\mathrm{MoS}_2/\mathrm{Mo}_2\mathrm{C}\)) & Chemical vapor carbonization & XRD, EDS, SEM, XPS & Solar simulator & 34,000 \(\mu\mathrm{mol\,g^{-1}\,h^{-1}}\) & 41.4 & Superactive co-catalyst & \cite{shao2019carbonized} \\
			\hline
			\(\mathrm{Ti}_3\mathrm{C}_2/\mathrm{ZnIn}_2\mathrm{S}_4\) & Solvothermal & XRD, TEM, FESEM, EDS, DRS, HRTEM, SEM & Visible (300 W Xenon lamp) & 978.7 \(\mu\mathrm{mol\,g^{-1}\,h^{-1}}\) & 24.2 & Binary composites & \cite{huang2022delaminating} \\
			\hline
			BN/MXene/\(\mathrm{ZnIn}_2\mathrm{S}_4\) & Calcination, etching, solvothermal & UPS, EPR, VB-XPS, Mott-Schottky & Visible & 1455 \(\mu\mathrm{mol\,g^{-1}\,h^{-1}}\) & – & S-scheme heterojunction & \cite{dai2023rational} \\
			\hline
			1T-\(\mathrm{MoS}_2/\mathrm{Ti}_3\mathrm{C}_2/\mathrm{TiO}_2\) & Hydrothermal & XRD, SEM, XPS, FESEM, BET, DRS & Sunlight (300 W Xe lamp) & 9738 \(\mu\mathrm{mol\,g^{-1}\,h^{-1}}\) & – & Synergistic charge transfer & \cite{li20191t} \\
			\hline
			\(\mathrm{Ti}_3\mathrm{C}_2/\mathrm{ZnO}\) & Electrostatic self-assembling & XRD, TEM, UV-Vis & Visible light & 4156 \(\mu\mathrm{mol\,g^{-1}\,h^{-1}}\) & – & Semiconductor-based heterojunctions & \cite{irfan2022construction} \\
			\hline
			\(\mathrm{WO}_3/\mathrm{Ti}_3\mathrm{C}_2/\mathrm{ZnIn}_2\mathrm{S}_4\) (WTZ) & Anaerobic solvothermal & XRD, TEM, HR-TEM, XPS, SEM, UV-Vis & Visible light & 7390 \(\mu\mathrm{mol\,g^{-1}\,h^{-1}}\) & – & Z-scheme heterojunction & \cite{wu2025constructed} \\
			\hline
			\(\mathrm{TiO}_2/\mathrm{Fe}_2\mathrm{O}_3@\mathrm{Ti}_3\mathrm{C}_2\) & Hydrothermal & XRD, TEM, EIS, HR-TEM, XPS, SEM, UV-Vis & 300 W Xenon lamp & 1634.64 \(\mu\mathrm{mol\,g^{-1}\,h^{-1}}\) & – & Heterojunction composites & \cite{yao2023one} \\
			\hline
			\(\mathrm{TiO}_2/\mathrm{Ti}_3\mathrm{C}_2/\mathrm{g}\text{-}\mathrm{C}_3\mathrm{N}_4\) (TTC) & Calcination & XRD, TEM, HRTEM, SEM, XPS & Xe 300 W lamp & 1150 \(\mu\mathrm{mol\,g^{-1}\,h^{-1}}\) & – & Heterostructure system & \cite{hieu2021tio2} \\
			\hline
		\end{tabular}
	\end{table}
	
	\subsection{Z-Scheme Photocatalysts}
\begin{figure}[htbp]
		\centering
		\fbox{\includegraphics[width=\linewidth]{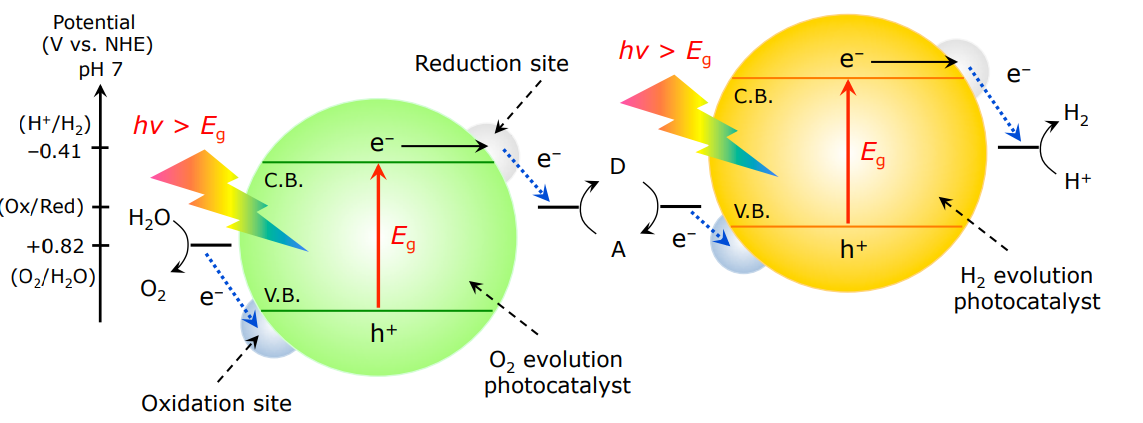}}
		\caption{Traditional(Liquid) Z Scheme. Adapted with permission from Ref.~\cite{maeda2010photocatalytic}, \copyright \,Copyright (2010) American
			Chemical Society.}
		\label{ZScheme}
	\end{figure}

	It is one of the common schemes as shown in Fig.~\ref{ZScheme}, representing a photocatalytic charge-transfer mechanism that mimics natural photosynthesis. In this mechanism, the water-splitting reaction proceeds in two sequential stages—one facilitating hydrogen (\(\mathrm{H_2}\)) evolution and the other enabling oxygen (\(\mathrm{O_2}\)) evolution. These two half-reactions are interconnected through a shuttle redox couple (Red/Ox) present in the solution. This process typically involves two distinct semiconductors (A and B) that absorb light and operate synergistically to enhance charge separation and improve the overall efficiency of the redox reactions.
	
	A typical Z-scheme photocatalytic system consists of two semiconductors, as depicted in Fig.~\ref{ZScheme}, commonly referred to as Photocatalyst A and Photocatalyst B. Examples include graphitic carbon nitride (\(\mathrm{g\text{-}C_3N_4}\)) as Photocatalyst A and titanium dioxide (\(\mathrm{TiO_2}\)) as Photocatalyst B. In advanced configurations, MXenes such as \(\mathrm{Ti_3C_2T_x}\) have also been integrated as Photocatalyst B due to their exceptional properties. Upon light irradiation, both semiconductors absorb photons, resulting in the excitation of electrons from their valence bands (VB) to conduction bands (CB), leaving behind holes in the VB. An electron mediator---either a redox couple like \(\mathrm{I^-}/\mathrm{I_3^-}\) or \(\mathrm{Fe^{3+}}/\mathrm{Fe^{2+}}\), or a solid conductor such as reduced graphene oxide (rGO) or MXenes---facilitates charge transfer between the two photocatalysts. In this Z-scheme mechanism, the photogenerated electron in the CB of Photocatalyst B recombines with the hole in the VB of Photocatalyst A through the mediator. This selective recombination enables effective charge separation, leaving strong oxidizing holes in the VB of Photocatalyst B and strong reducing electrons in the CB of Photocatalyst A. These charge carriers then drive redox reactions: electrons reduce protons to generate hydrogen (\(\mathrm{H_2}\)), while holes oxidize water to produce oxygen (\(\mathrm{O_2}\)). MXenes like \(\mathrm{Ti_3C_2T_x}\) are particularly attractive as electron mediators due to their high electrical conductivity.
	
	Implementing a Z-scheme mechanism lowers the energy demand for each photocatalytic step, thereby broadening the range of usable wavelengths. In this approach, H$_2$ evolution takes place at approximately 660\,nm, while O$_2$ evolution occurs around 600\,nm. This represents a significant extension compared to traditional water-splitting systems, which rely on single-step photoexcitation in a single semiconductor and are limited to shorter wavelengths ($\sim$460\,nm).
	More details on Z-scheme water splitting, including its historical development and advancements in material design, can be found in various review papers~\cite{maeda2013z,wang2018mimicking}.
	
	The efficiency of the Z-scheme mechanism largely depends on the redox potential of the mediator and its interaction with the semiconductor surface. For further insights, comprehensive reviews are available in \cite{wang2018mimicking,maeda2013z}.

	The latest scheme for studying photocatalysts, following the Z-scheme, includes Type-I and Type-II Heterojunctions, advanced S-scheme heterojunctions\cite{li2016z,zhang2022emerging,hasham2024enhanced,xu2020unique}, direct Z-schemes\cite{wang2020direct,low2017review,xu2018direct,li2021review}, and dual Z-schemes\cite{malefane2023understanding,thambiliyagodage2022fabrication}. These newer mechanisms improve charge separation and transfer efficiency, crucial for applications like water splitting.
	
	MXene-based Z-scheme systems facilitate efficient charge transfer by mimicking natural photosynthesis\cite{ashraf2024recent,zhuang2024organic,li2021chlorophyll,mishra2024mxene}.
	
	\subsection{Ternary Heterostructure(All-Solid-state Z-Scheme)}
	
	In photocatalysis, a ternary heterostructure, as shown in Fig.~\ref{milestone}, involves the integration of three different materials—usually semiconductors—into a single composite system to enhance photocatalytic performance. This configuration is designed to maximize light harvesting across a broader spectral range and facilitate efficient charge separation and transfer between interfaces. The strategic combination of materials allows for the formation of multiple heterojunctions, which help in minimizing charge recombination and promoting directional charge migration. Ternary heterostructures can be engineered in various configurations such as Z-schemes, S-schemes, or type-II band alignments, depending on the desired charge transfer pathway. When incorporated with advanced materials like MXenes (instead of noble metals like \( \mathrm{Pt} \) and \( \mathrm{Au} \), which have large work functions), which can act as conductive bridges or active co-catalysts, these systems exhibit superior photocatalytic activity for applications including hydrogen evolution, pollutant degradation, and \( \mathrm{CO}_2 \) reduction.
	
	Ranjith \textit{et al.}~\cite{ranjith2024interfacial}, as discussed earlier, employed this scheme to successfully enhance light absorption and charge separation, significantly improving photocatalytic \( \mathrm{H}_2 \) production. This approach offers a promising strategy for efficient, solar-driven hydrogen generation using MXene-based heterostructures.

	\subsection{S-Scheme Heterojunction (Step-Scheme)}
	
\begin{figure}[htbp]
		\centering
		\fbox{\includegraphics[width=\linewidth]{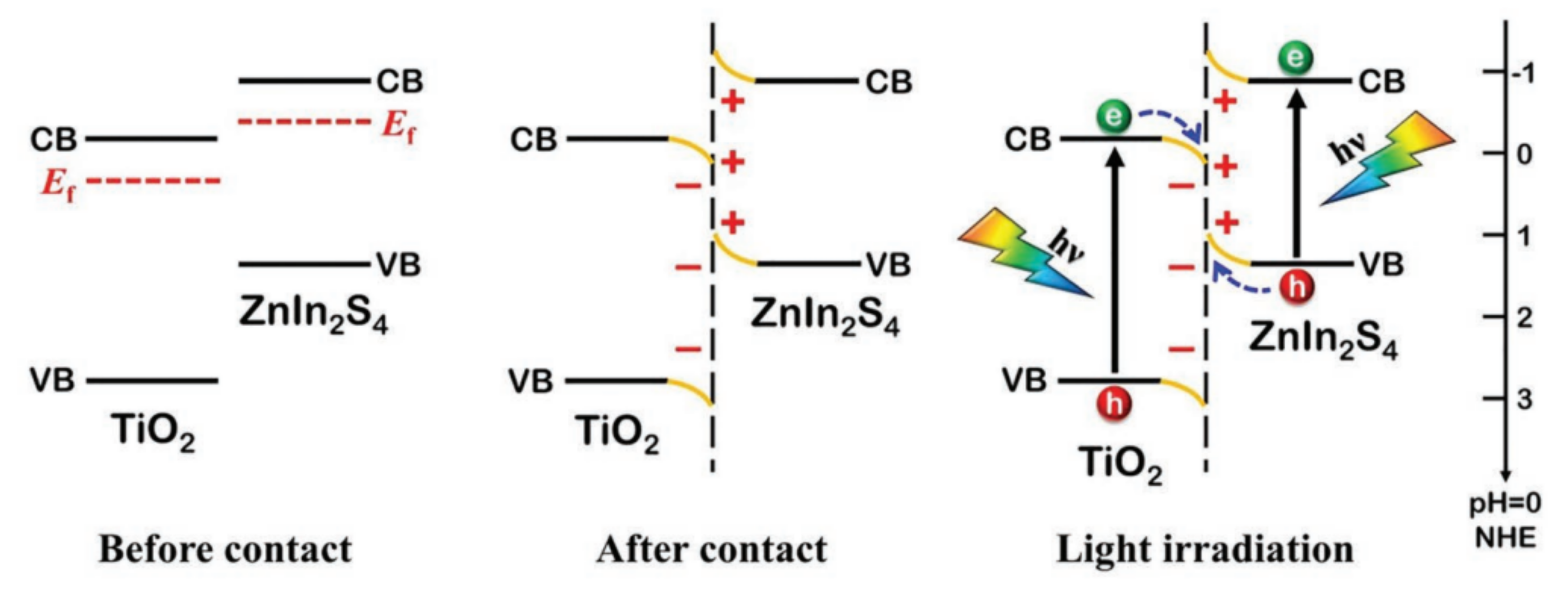}}
		\caption{S-Scheme. Adapted with permission from Ref.~\cite{zhang2022emerging}, \copyright \, (2022) Wiley Online Library.}
		\label{SScheme}
	\end{figure}

	J. Fu \textit{et al}.~\cite{fu2019ultrathin} recently proposed S-scheme heterojunction photocatalytic system that aims to combine the benefits of Type-II and Z-scheme systems while mitigating their drawbacks\cite{xu2020s,zhang2022emerging,bai2022assembling}. In this configuration, two semiconductors with staggered band positions form a solid-state interface that creates an internal electric field due to Fermi level equilibration and band bending. Upon light irradiation, both semiconductors generate electron-hole pairs. Selective recombination takes place between the photogenerated electrons in the conduction band (CB) of the low-energy photocatalyst and the holes in the valence band (VB) of the high-energy photocatalyst, facilitated by the built-in electric field, as illustrated in Fig.~\ref{SScheme}.
	This selective recombination preserves the electrons in the CB of the high-energy photocatalyst and the holes in the VB of the low-energy photocatalyst—thus retaining strong redox potentials(sufficiently positive or negative energy levels i.e valence and conduction band positions that are energetically capable of driving redox reactions). The electrons participate in reduction reactions (e.g., \( \mathrm{H}^+ \rightarrow \mathrm{H}_2 \)) and the holes in oxidation reactions (e.g., \( \mathrm{H}_2 \mathrm{O} \rightarrow \mathrm{O}_2 \) or pollutant degradation).
	Notably, S-scheme systems do not require external mediators and achieve enhanced charge separation and photocatalytic activity due to the synergistic effect of internal electric field, band bending, and coulombic interactions. Its mechanism has been beautifully explained in literature\cite{xu2022design}. The S-scheme heterojunction not only enhances the separation and transport of charge carriers but also optimizes the redox potential of the resulting heterojunction system.
	
	MXenes, such as \( \mathrm{Ti}_3 \mathrm{C}_2 \mathrm{T}_x \), are often used in S-scheme systems as either active photocatalysts or conductive bridges, owing to their high electrical conductivity, large surface area, and tunable surface functionalities.
	
	\section{Challenges and future perspective}
	Currently, commercially available \( \mathrm{H}_2 \) is primarily produced from fossil fuels such as coal (black \( \mathrm{H}_2 \)) and natural gas (grey \( \mathrm{H}_2 \)). However, the production of \( \mathrm{H}_2 \) from fossil fuels inevitably results in \( \mathrm{CO}_2 \) emissions.
	
	Photocatalysis is the urgent and important need of an hour.
	Achieving 100\% quantum efficiency in photocatalytic overall water splitting remains a significant challenge in photocatalysis research. It requires meeting two fundamental criteria\cite{takata2020photocatalytic} as follows:

	\begin{enumerate}
		\item {Efficient Charge Transport:} Photoexcited charge carriers must effectively migrate to the surface reaction sites before undergoing recombination within the bulk material.
		\item {Sequential Electron and Hole Transfer:} The hydrogen evolution reaction (HER) necessitates the injection of two electrons, while the oxygen evolution reaction (OER) requires the transfer of four holes. Both processes must occur in succession without any undesired backward charge transfer.
	\end{enumerate}
	
	However, the high probability of backward electron transfer makes it exceedingly difficult to achieve an external quantum efficiency (EQE) exceeding 50\%, even with ultraviolet-responsive photocatalysts\cite{,li2014proposed,takata2020photocatalytic,li2019photocatalytic,zou2001direct,chiang2018efficient}. Therefore, a key challenge in photocatalysis is to determine whether a 100\% efficiency can be realized by completely preventing backward electron transfer and optimizing the photocatalyst structure.
	
	Photocatalysts such as \( \mathrm{SrTiO}_3\!:\mathrm{Al} \), when modified with \( \mathrm{Rh} \), \( \mathrm{Cr}_2\mathrm{O}_3 \), and \( \mathrm{CoOOH} \), have exhibited remarkable photocatalytic performance, with an EQE of 95.9\% at 360 nm \cite{takata2020photocatalytic}.
	However, their optical absorption is largely limited to the ultraviolet region, which restricts solar energy utilization, and the solar-to-hydrogen (STH) energy conversion efficiencies obtained in the study remain around 0.76\%, which is approximately an order of magnitude lower than the 10\% threshold necessary for practical implementation.
	
	In contrast, MXene-based photocatalysts, because of their broadband light absorption that extends into the visible range, offer a significant advantage for enhancing solar-to-hydrogen (STH) efficiency. However, several challenges remain as follows:
	
	\emph{Structural Stability}: MXenes are prone to oxidation, which limits their long-term stability in aqueous environments. The oxidation and degradation of MXene in aqueous environments pose significant stability challenges, requiring advanced protective strategies.
	
	\emph{Charge Recombination}: High electron conductivity can lead to undesired charge recombination, reducing photocatalytic efficiency.  Cocatalysts, doping, and defect engineering are essential strategies to mitigate this issue.
	
	\emph{Scalability and Cost}: Developing cost-effective synthesis methods and scalable processes is a key challenge for large-scale MXene-based photocatalyst applications.
	
	\emph{Bandgap Tuning Issues}: Optimizing band alignment is crucial for enhancing photocatalytic efficiency. Various strategies, including heterostructure engineering and chemical modifications, are under investigation.
	
	Future research could significantly improve the photocatalytic efficiency of MXene-based materials for hydrogen production by exploring the following directions:
	
	\begin{itemize}
		\item Construct MXene-based heterostructures, such as S-scheme systems with \( \mathrm{MoS}_2 \) or \( \mathrm{g\text{-}C}_3\mathrm{N}_4 \), to facilitate more effective charge separation and broaden light absorption.
		\item Adjust electronic properties by modifying surface terminations (e.g. –O, –OH, –F) or incorporating dopants to improve overall photocatalytic activity.
		\item Investigate the phase stability of MXenes to strengthen their resistance to corrosion and extend their operational life.
		\item Introduce engineered defects and cocatalysts such as \(\mathrm{Pt}\) or \(\mathrm{RuO_2}\) to enhance the mobility of the charge carrier and promote efficient hydrogen evolution.
		\item Refine the design of S-scheme photocatalysts to achieve stronger redox capabilities and minimize electron–hole recombination.
		\item Focus on scalable synthesis techniques, device integration strategies, and holistic assessments of environmental and economic feasibility.
	\end{itemize}
	
	\section{Conclusion}
	MXene-based photocatalysts hold significant potential for efficient hydrogen production owing to their excellent electrical conductivity, tunable surface chemistry, and favorable band structures. However, several challenges remain, including controlling surface terminations, improving stability under photocatalytic conditions, and achieving efficient integration with other semiconductor materials. Future progress will rely on innovative synthesis methods to produce high-quality MXenes with tailored properties, the rational design of heterostructures that optimize light absorption and charge carrier dynamics, and the development of advanced charge separation strategies to minimize electron–hole recombination. Continued research in these areas will be crucial in unlocking the full potential of MXene-based systems for solar-driven water splitting and sustainable hydrogen generation.
	
	\section{Acknowledgment}
	
	\textbf{DPR} acknowledges Anusandhan National Research Foundation (ANRF), Govt. of India, through Sanction Order No.:CRG/2023/000310, \& dated:10 October, 2024.\\
	
	A. Laref acknowledges support from the "Research Center of the Female Scientific and Medical Colleges",  Deanship of Scientific Research, King Saud University.\\
	The research is partially funded by the Ministry of Science and Higher Education of the Russian Federation as part of the World-Class Research Center program: Advanced Digital Technologies (contract No. 075-15-2022-312 dated 20.04.2022).
	
	\section*{Author contributions}
	\begin{itemize}
		\item \textbf{C. B. Subba:} First author, Formal analysis, Visualization, Validation, survey of literature, Writing-original draft, writing-review \& editing.
		\item \textbf{Prasad Mattipally}:Formal analysis, Visualization, Validation, writing-review \& editing. 
		\item \textbf{J. Sivakumar}:Formal analysis, Visualization, Validation, writing-review \& editing. 
		\item \textbf{A. Laref}:Formal analysis, Visualization, Validation, writing-review \& editing. 
		\item \textbf{A. Yvaz}:Formal analysis, Visualization, Validation, writing-review \& editing. 
		\item \textbf{D. P. Rai:} Project management, Supervision, Resources, Formal analysis, Visualization, Validation, writing-review \& editing. 
		\item \textbf{Z. Pachuau}:Formal analysis, Visualization, Validation, writing-review \& editing. 
	\end{itemize}
	
	\section*{Data Availability Statement}
	Data is available in the article.

	
	\bibliography{sn-bibliography}
	
\end{document}